\documentclass[letterpaper,twocolumn,10pt]{article}
\usepackage{usenix-2020-09}

\usepackage[colorinlistoftodos,prependcaption,textwidth=1.22cm]{todonotes}
\marginparwidth=\dimexpr \marginparwidth + 1cm\relax
\usepackage{pifont}
\usepackage{wasysym, multirow}
\usepackage{graphicx}
\usepackage{booktabs}
\usepackage{float}
\usepackage{soul}
\usepackage{xcolor}
\usepackage{microtype}
\usepackage{times}
\usepackage{bookmark}

\newcommand{\market}{\texttt{IMPaaS.ru}}
\newcommand{\impaas}{\texttt{IMPaaS}}

\newcommand{\crypto}{\texttt{Crypto}}
\newcommand{\money}{\texttt{MoneyTransfer}}
\newcommand{\commerce}{\texttt{Commerce}}
\newcommand{\social}{\texttt{Social}}
\newcommand{\services}{\texttt{Services}}
\newcommand{\other}{\texttt{Other}}

\newcommand{\LIM}[1]{\texttt{\textbf{Ch{#1}}}}

\usepackage{tikz}

\begin{document}

\date{}

\title{\Large \bf Know Your Cybercriminal: Evaluating Attacker Preferences by Measuring Profile Sales on an Active, Leading Criminal Market for User Impersonation at Scale}

\author{
{\rm Michele Campobasso}\\
m.campobasso@tue.nl\\
Eindhoven University of Technology
\and
{\rm Luca Allodi}\\
l.allodi@tue.nl\\
Eindhoven University of Technology
} 

\maketitle





\begin{abstract}
In this paper we exploit market features proper of a leading Russian cybercrime market for user impersonation at scale to evaluate attacker preferences when purchasing stolen user profiles, and the overall economic activity of the market. 
We run our data collection over a period of $161$ days and collect data on a sample of $1'193$ sold user profiles out of $11'357$ advertised products in that period and their characteristics.
We estimate a market trade volume of up to approximately $700$ profiles per day, corresponding to estimated daily sales of up to $4'000$ USD and an overall market revenue within the observation period between $540k$ and $715k$ USD. We find profile provision to be rather stable over time and mainly focused on European profiles, whereas actual profile acquisition varies significantly depending on other profile characteristics. Attackers' interests focus disproportionally on profiles of certain types, including those originating in North America and featuring \crypto\ resources.  We model and evaluate the relative importance of different profile characteristics in the final decision of an attacker to purchase a profile, and discuss implications for defenses and risk evaluation.

  
\end{abstract}

\section{Introduction}
\label{sec:intro}

Studying underground communities can provide important insights into cybercriminal actions and threat levels~\cite{allodi2017economic,soska2015measuring,aliapoulios2021swiped}. In particular, the evaluation of underground markets can help quantifying the risk on final users posed by cybercriminal activities. For example, the observation of criminal ecosystems has been employed in research to identify innovative or emergent threats, and the monitoring of trade activity to evaluate their associated impact on final users~\cite{bhalerao2019mapping, campobasso2020impersonation,aliapoulios2021swiped}. On the other hand, obtaining reliable data from criminal marketplaces is an increasingly challenging activity~\cite{turk2020tight} as platform administrators start deploying anti-crawling measures~\cite{campobasso2022threat} and access control measures vetting accounts requesting access to their community(-ies)~\cite{allodi2017economic}. Furthermore, data collected in these underground places is often censored or missing, for example due to infrastructural failures at certain crawling times. This is particularly challenging for longitudinal studies (of any length) aiming at monitoring market/community evolution over a period of time, measuring differences in outcomes or, for example, product provision~\cite{soska2015measuring}. Data is hard to interpret as well, as generally only indirect signals of events are available for inference (e.g., user feedback as a proxy variable for product sales). Exceptions exist for leaked databases, although this generally allows studying markets that have already died or collapsed, oftentimes as a result of the leak itself. In other words, the opportunity to reliably study threat levels posed by active underground markets, their relevance globally and over time, and the overall size of the underlying economy supporting those threats is rare. 





\subsection{Research gap and contribution} 
In this work, we study a unique data collection of sale volumes and trends on \market\ (pseudonym), a leading, invite-only Russian underground platform currently active and operating as the main provider for Impersonation-as-a-Service in the criminal underground~\cite{campobasso2020impersonation} to evaluate the overall threat levels it poses globally to the population of Internet users,  quantify the size of the underlying market economy supporting these attacks, and evaluate attacker preferences when choosing a profile to purchase. 
This paper's contribution is multi-fold:

\begin{enumerate}
    \item We present a thorough data collection methodology addressing the key challenges of monitoring the evolution of specific products in the market, while avoiding anti-crawler technologies and under the constraints introduced by monitoring closed-access marketplaces. We discuss the necessary trade-offs and present the respective solutions. 
    \item We devise a robust data analysis methodology addressing uncertainties in the data collection resulting from those trade-offs; the proposed methodology handles high dimensionality data while capturing all variance in the original variables and maintaining full transparency on the relation between dimensions and outcome;
    \item We provide an extensive analysis of the size and relevance of \impaas\ as a global threat model, estimating volumes of acquired profiles across regions, profile characteristics, and time, hence providing a realistic proxy measure of actual victimization rates.
    \item We provide a characterization of attackers' purchasing decisions and their price sensitivity across profile types. Whereas limited to the setting of \market, our characterization provides novel insights on criminal purchase decisions and associated trends;
    \item We provide a robust estimation of the revenues of the analyzed criminal market. We analyze sale trends and derive market economic size employing a mixture of real, predicted, and simulated sale data;
    \item We discuss our findings on attacker preferences 
    and their relation to attack surface evaluation, and to the identification of possible countermeasures in response to market observations.
    \item We share all the datasets and the crawling infrastructure at \url{https://security1.win.tue.nl}.
\end{enumerate} 

This manuscripts proceeds as follows: Sec.~\ref{subsec:related_work} discusses related work; Sec.~\ref{sec:methodology} breaks down the problem at hand and presents our methodology for data collection and analysis, whereas Sec.~\ref{sec:results} first presents an overview of the data, to then delve in sale activity in \market. Sec.~\ref{sec:discussion} discusses findings and concludes the paper.

\section{Background \& Related work}
\label{subsec:related_work}

\market\ specializes in offering \textit{user profiles} to attackers (i.e., \market's customers); user profiles are bundles of information stolen from victims across the globe via malware infection, allowing attackers to replicate a victim's browser environment with the purpose of bypassing web anti-fraud techniques such as risk-based authentication~\cite{campobasso2020impersonation, marriott_2021, woods_boddy_backer_2020, pascu_2021, gracey-mcminn_2021, wiefling2019really}. 
The user profiles traded on \market\ include stolen credentials and cookies of the victim browser, as well as additional information necessary to mimic the `appearance' (fingerprint) of the victim's browser to an authentication service~\cite{campobasso2020impersonation,wiefling2019really}.  
Customers of \market\ can browse across the portfolio of offered profiles and evaluate them by inspecting the list of websites for which stolen credentials are present, the country of origin of the profile, when the information was first harvested and last updated, etc.~\cite[for a full enumeration]{campobasso2020impersonation}. When buying a profile, the customer can download the bundle of information within that profile together with a Google Chrome browser extension developed by the \market\ operators. The purpose of the browser extension is to allow the customer to instrument their browser with the information contained in the purchased user profile, in order to replicate the victim's browsing environment; the replicated environment can then be used to conduct the impersonation attack. Importantly, upon purchase of a profile, the profile is unlisted from the market. On the one hand this assures a profile is purchased only once; on the other, it provides a method to precisely measure sales. 
Interestingly, recent work by Lin et al.~\cite{lin2022phish} proposes techniques to evade risk-based authentication (RBA) services similar to those originally introduced by \market\ (including stealing information from the victim's environment, and re-producing these in the attacker's by means of a browser extension), and find that authentication services are indeed vulnerable to these attacks. 
The threat posed by impersonation attacks against RBA demonstrated in \cite{lin2022phish} was first described in \cite{campobasso2020impersonation}, together with a description of the \impaas\ threat model, \market\ pricing model, and features of the traded product (i.e., the user profiles).
Differently from these works, in this paper we study attackers' profile purchasing behavior by monitoring patterns in product offering and sales from the market activity itself, to derive insights on attackers' decisions when selecting targets to impersonate under the \impaas\ model. Further, by analyzing actual sales data from \market, we evaluate the overall relevance of the \impaas\ threat worldwide.

To contextualize this work, we refer to the cyber risk model proposed by Woods and Böhme~\cite{woods2021sok} (depicted below, in gray).

\begin{figure}[!htbp]
    \centering
    \vspace{-3mm}
    \includegraphics[width=0.91\columnwidth,keepaspectratio]{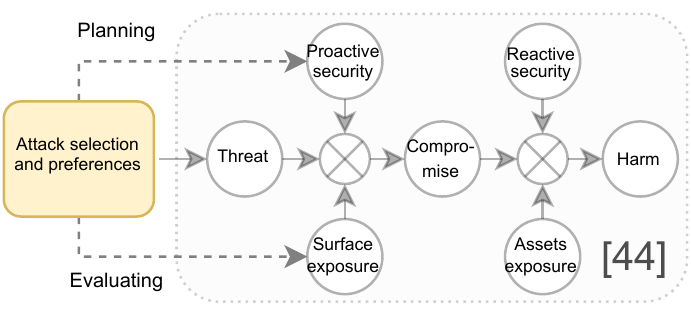}
    \vspace{-4mm}
\end{figure}

\noindent
The risk model presented in \cite{woods2021sok} identifies a number of latent variables whose interplay characterizes the overall risk picture, starting from `Threat' and leading to `Harm'. On the other hand, threats do not materialize `out of thin air'; rather, they are generated by (human) attackers that, whether through access to the criminal ecosystem or by their (or their organization's) own means, consciously choose their targets and suitable attack technologies or methods~\cite{collier2019booting,allodi2017economic,herrprimer}. Critically, being able to characterize attacker preferences before the threat materializes can help defenders in better devising their `proactive security', and can provide insights on the actual exposure of an organization to said threats. To capture this, we propose to extend Woods and Böhme's model by including `\textit{Attack selection and preferences}' as a precursor step to the arrival of a `Threat'. Specifically, by studying \market\ sales, in this paper we reconstruct the attacker preferences leading to the actualization of the \impaas\ threat, and discuss implications on defenses and attack exposure.

\vspace{-0.1in}
\subsubsection*{Related Work}
Gathering data to study cybercriminal ventures is a longstanding problem. Often, data comes from manual collection~\cite{benjamin2016conducting}, incomplete or partial crawling~\cite{portnoff2017tools}, or relatively outdated leaks of underground marketplaces~\cite{van2020go, benjamin2016conducting, motoyama2011analysis, portnoff2017tools, overdorf2018under, bhalerao2019mapping}. The objective difficulty linked with the collection and analysis of this type of data results in multiple studies looking at the same or similar (e.g., updated) data~\cite{van2020go, akyazi2021measuring, bhalerao2019mapping, van2018plug}. Several authors develop specialized crawlers to scrape the target infrastructure~\cite{soska2015measuring, zhang2019key}, produce tools capable of obtaining fresh data over time across underground communities~\cite{pastrana2018crimebb} and tackle the problem of developing general crawlers flexible enough to target multiple criminal forums or marketplaces~\cite{jiang2012focus, campobasso2022threat}; some of these solutions propose anti-crawler detection techniques to avoid detection from the administrators of the crawled communities~\cite{pastrana2018crimebb, portnoff2017tools, soska2015measuring, campobasso2022threat}. 

Aside from the data collection, the analysis of this type of data presents foundational challenges: the processes behind its generation are oftentimes at least partially unknown~\cite{anderson2013measuring}, and estimates (particularly of an economic nature regarding sales and purchase activity) can only be approximated~\cite{soska2015measuring,van2018plug}. \textit{Post-mortem} analyses of cybercriminal revenues based on data provided from law enforcement following takedowns of markets~\cite{noroozian2019platforms, laarschot2021risky} or leaked data~\cite{aliapoulios2021swiped, brunt2017booted, karami2013understanding} are often among the most accurate estimates one can derive, albeit generally on criminal marketplaces or communities that no longer exist. The difficulty of this data collection and analysis process sometimes results in contrasting and/or disputed estimates~\cite{mcafee2014net, by2013norton}; \cite{anderson2013measuring} provides an additional commentary. 
Live data collection with a clear data generating process aiding its analysis is rare in criminal settings, albeit crucial to obtain reliable estimates of still-alive and evolving cybercriminal activities, and to develop tailored countermeasures to operating threats~\cite{armin20152020}. 
\section{Methodology} 
\label{sec:methodology}
As part of our approach, we first identify critical aspects of the data collection and the problem at hand. These challenges are posed by the nature of the data and of the problem we address; 
therefore, it is useful to detail these challenges upfront. Tab.~\ref{tab:limitations} shows which methodological steps address them. 

\begin{figure*}[t]
    \centering
    \includegraphics[width=0.7\textwidth,keepaspectratio]{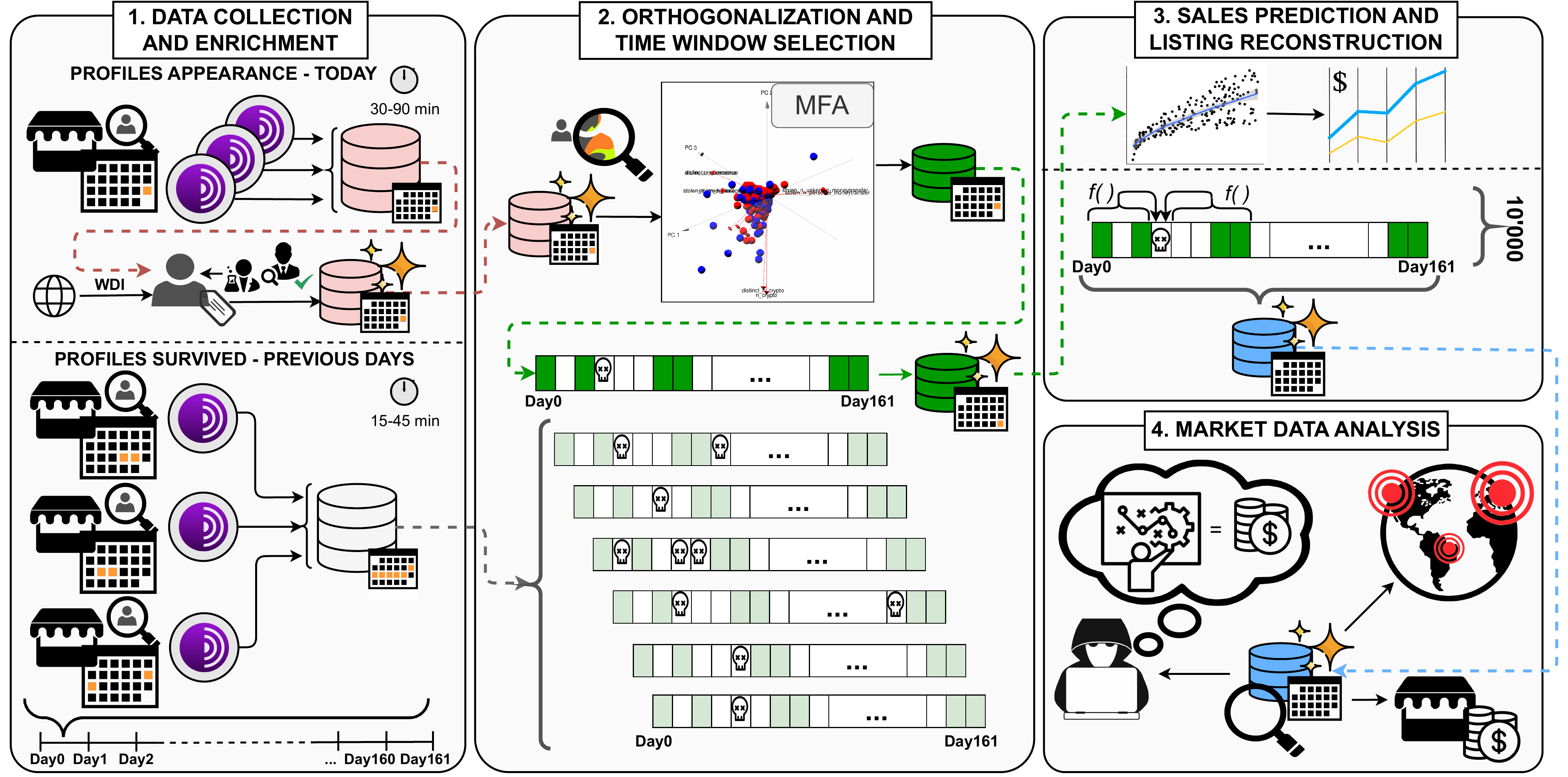}
    \caption{Methodology overview. Acronyms: WDI = World Development Indicator; MFA = Multiple Factor Analysis.}
    \vspace{-4mm}
    \label{fig:methodology}
\end{figure*}

\begin{table}[t]
\centering
\caption{Relationship between challenges and mitigating step(s) of the methodology.}
\vspace{2mm}
\label{tab:limitations}
\small
\begin{tabular}{p{0.01\columnwidth}p{0.31\columnwidth}p{0.045\columnwidth}p{0.045\columnwidth}p{0.045\columnwidth}p{0.045\columnwidth}p{0.045\columnwidth}p{0.045\columnwidth}}
\toprule
                                            & & \LIM{1} & \LIM{2} & \LIM{3} & \LIM{4} & \LIM{5} & \LIM{6} \\
\midrule
\multirow{5}{*}{\rotatebox{90}{Method. step}}  & Data collection          & $\times$ & $\times$ & $\times$ &  &  & \\
                                    & Data enrichment           & & & & & & \\
                                    & Feat extract \& orthog       & & & & & $\times$ & \\
                                    & Data diagonalization         & & & & & & $\times$ \\
                                    & Sales pred \& sim.      & $\times$ & $\times$ & & $\times$ & $\times$ & \\
\bottomrule
\end{tabular}
\end{table}


\subsection{Challenges}
\label{subsec:problem_def}

\smallskip
\noindent\LIM{1}. \textbf{Reliability of criminal infrastructures.} Connectivity to criminal infrastructures (\market\ included), critical for prolonged crawling activities monitoring market evolution such as the one performed for this research, is often unreliable. 

\smallskip
\noindent\LIM{2}. \textbf{Bandwidth of TOR network.} To minimize exposure of the crawling activity, it should be performed over TOR. It is critical for the data collection to use as little bandwidth as possible not to compromise other TOR users' experience. 

\smallskip
\noindent\LIM{3}. \textbf{Crawling prevention measures.} Prior work showed that \market\ employs anti-crawler measures that can lead to user banning; as obtaining \market\ access can require up to $\approx$ 1 month, it is critical that the crawler accounts for the countermeasures in place.

\smallskip
\noindent\LIM{4}. \textbf{Repeated measurements.} To monitor product evolution on \market\ we must monitor their (dis)appearance as time progresses. This requires repeated (re-)measurements of the platform at different moments in time and within sufficiently small time windows. 
However, these time windows cannot be too small due to the risks connected to \LIM{3}, meaning that a trade-off exists between sampling completeness and persistence of market access. 

\smallskip
\noindent\LIM{5}. \textbf{Measurement of aggregate, high-dimensionality effects.} Uncovering the decision process of attackers operating on \market\ to acquire a profile requires transparently linking highly-dimensional data~\cite{campobasso2020impersonation} with sale observations while preserving as much as possible (or all) of the original variance in the observations. Further, only being able to observe the aggregate effect of customers purchase decisions increases uncertainty in the model. 

\smallskip
\noindent\LIM{6}. \textbf{Accurately measuring sales.} We must distinguish a profile `disappearance' during a crawling session caused by its `sale' rather than by temporary glitches or effects.  

\subsection{Methodology steps}\label{sec:methodology-steps}

We devise a multi-stage methodology for collecting and enriching \market\ data, with the aim of modelling market sales and gaining quantitative insights into the market economy and customer purchase decisions. 
Fig.~\ref{fig:methodology} gives a bird-eye of our methodology. 

\subsubsection{Data collection}
\label{subsubsec:meth_data_collect}
To conduct our study, we exploit the fact that \market\ only lists \textit{still available} profiles on their listing and removes items only through a sale, or a reservation (as verified by us, the reservation mechanism allows a customer to reserve a profile for $30$ minutes, temporarily removing the product from the listing).
We exploit this mechanism to collect data on profile appearances and their persistence on the market. 
From the first data collected, we notice that the chances of sale for a profile sharply decrease after the first day and become negligible after the sixth day. Therefore, we establish six days as the time window of choice during which to monitor a profile after its appearance on \market. 
To scrape the market while addressing \LIM{3}, we create six crawler instances: three \textit{appearance} and three \textit{persistence} crawlers. This choice was made after considerable trial and error (leading to account banning on the market) to strike a balance between the volume of data to collect, the level of stealth needed to remain `under the radar', and the number of available accounts we could `burn through' during the data collection. This last requirement is particularly critical as accounts were not easy to obtain across other platforms and communities. The \textit{appearance} crawlers reach the market's listing section at midnight (Moscow time), tasked with obtaining a full description of appeared profiles in the previous $24$ hours relative to the start of the crawling on day $d$. We decided to crawl during the eastern-Europe night to reduce our impact on the platform’s responsiveness during (likely) active hours, and therefore reduce possible alerts triggering further investigation from the market admins. In mitigation to \LIM{2,3}, and to keep the architecture as simple as possible, the three crawlers split the workload independently by collecting the full list of appeared profiles and selecting only the $1/3$ that corresponds to their crawler id.
Within their $1/3$, each crawler randomly selects 25\% of the listed products. Initially, we attempted to download the whole offer of the day, but crawling sessions often exceeded six hours, which would in turn introduce large inconsistencies in the temporal dimensions of the measured sales. 
This procedure allows us to limit data crawling visibility (\LIM{3}) while collecting a representative and valid sample of data on profile appearance and characteristics. 
For each appeared user profile we collect the full set of features that characterize it. The result is a data collection that fully represents what the market customer sees when viewing an item. Additionally, one \textit{appearance} crawler is tasked with collecting a recap, offered by \market, of the number of appeared profiles during the last $24$ hours. For each day $d$ in the \textit{observation} period up to day $D$, we aim at obtaining a data collection $L^d_0,\ \forall d \in [0..D]$ of all appeared user profiles on that day.
In parallel, the three \textit{persistence} crawlers monitor the market to collect the names of the profiles still available. The \textit{persistence} crawlers monitor appeared profiles in each day $d$ for six consecutive days since $d$. 
Each \textit{persistence} crawler is assigned a period of two days relative to the (midnight of) the day in which the crawler is run. The \textit{persistence} crawlers collectively generate, for each day $d$, a dataset $L^d_{1..6}$ containing the IDs of the appeared user profiles on day $d$ (and not yet sold) across each \textit{monitoring} day $n\in[1..6]$ relative to $d$.
To limit the impact of \LIM{4} we probe the market for changes in product offering every $24$ hrs.

Each run of the three appearance crawlers requires $30-90$ minutes depending on the products offered, market responsiveness (\LIM{1}), and available bandwidth (\LIM{2}). The three persistence crawlers take $15-45$ minutes on average. We consider this sufficiently fast to mitigate \LIM{4}, while not aggravating \LIM{1-3}. We further mitigate \LIM{3} by throttling traffic. 

We implement the crawlers using instrumented TOR Browser\cite{torbrowser} instances via the Selenium\cite{selenium}-based 
library \textit{tbselenium}\cite{tbselenium} to generate traffic from an instrumented browser without having to tinker with technical details that may raise a red-flag in crawler detection systems~\cite{campobasso2022threat}. Each crawler instance accesses a completely different TOR circuit to avoid using the same bastion host.   
Further, each of the crawler instances is assigned to a different user account under our control, limiting the activity of each account overall (\LIM{3}). Finally, to assure an as-complete-as-possible data collection in presence of \LIM{1} and \LIM{2}, the crawlers are designed to automatically adjust timeouts to refresh pages when those cannot be fetched on the first attempt, by doubling the default fetch timeout of $15$ seconds until the page is not successfully loaded, or retrying every $5$ minutes if the market is not reachable. The crawler keeps attempting to connect to the market until 2am Moscow Time. This choice is to limit noise in the data collection whereby profiles disappear before they are collected by our crawler (ref. \LIM{4}); this assures that comparisons across snapshots on different days remain meaningful. 



\label{subsubsec:data-enrichment}
We enrich obtained user profiles with data on the 2020 per capita GDP of the respective country of origin. To better reason about the characteristics of a profile we follow~\cite{campobasso2020impersonation}, and aggregate and classify available resources (stolen credentials originating from a specific website) for that profile in six categories: \services\ (delivery of physical or digital goods, such as Netflix or Gmail); \money\ (traditional payment, like PayPal or American Express); \crypto\ (payments via cryptocurrency circuits, such as Crypto or Bitpanda); \social\  (user-generated content, like Facebook or Twitter); \commerce\ (purchase or book goods from one or multiple vendors, like Amazon); \other\  (for otherwise non-classified resources). 
The classification is done manually by an author and independently checked for a random sample of 100 resources in a blinded process by a second author until conflicts are resolved. 

\subsubsection{Feature extraction and orthogonalization}
\label{subsubsec:meth_feature_extract}

Due to the high uncertainty inherently involved in reconstructing purchase decisions, let alone criminal ones, we employ a set of techniques to maximize the amount of information available to our modelling. The objective is to transform the data to prevent correlated, high-variance variables~\cite{campobasso2020impersonation} to dominate the resulting analysis, while not losing information in the transformation. That is also challenging because profile characteristics are naturally `nested' within groups of semantically related information on that profile~\cite{campobasso2020impersonation}: for example, both the available cookies and the available browser environments describe features of available browsers; 
as such, these features should \textit{not} be treated as independent entities. 
To accommodate for this we employ \textit{Multiple Factor Analysis} (MFA) as the method of choice to derive linearly uncorrelated dimensions of the data for our analysis~\cite{mfa}. MFA integrates Principal Component Analysis (PCA) for the numerical variables and Multiple Correspondence Analysis (MCA) for the categorical variables while preserving effects at the group level. 
As a result, the original variables collected in our dataset are projected over several, orthogonal `dimensions' with near-zero correlation, thus maximizing the explanatory power of each added dimension by removing overlap, helping in the identification of patterns in data and mitigating \LIM{5}. 


\paragraph{Data diagonalization and time window selection.}
\label{subsubsec:meth_data_diag}


The data diagonalization has the primary function of allowing us to make a well-informed decision on how wide the time window we consider, across the six days monitoring period, should be. A discussion of why this is necessary for modelling consistency and preserving the internal validity of this study is provided in Sec.~\ref{subsubsec:meth_pred-and-simulation}.
To inform this decision, we (a) estimate how many profiles are sold for each of the six monitoring days, and (b) evaluate whether sold profiles remain similar regardless of the day on which they are sold. 
To achieve (a), for each day, we mark a profile as sold if the product disappears after $n$ days and does not appear on any subsequent day. To do this we only keep records of days that we have fully monitored up to a certain monitoring day $n$ 
(i.e, $\bigcup_{d\in D} L^d_{0\ldots n}= \bigcup_{d\in D} L^d_{0} \cap L^d_{1} \cap \ldots \cap L^d_{n},\ $with$\ n\in[1..6]$). 
For example, if we collect $L^{d'}_0$ and $L^{d'}_1$ but not $L^{d'}_2$, we will keep $d'$ in $\bigcup_{d\in D} L^{d}_{0..1}$, but not in $\bigcup_{ d\in D}L^d_{0..2}$ (i.e., day $d'$ will result as a missing day in the diagonalized data for $n=2$). We then achieve (b) by simply comparing profile characteristics (orthogonalized via MFA) across profiles sold on different days. 
To distinguish `sold' from `reserved' profiles (\LIM{6}), we check for every collection $L^d_0$ if a profile disappeared in any of $L^d_{1..n}, n\in[1..6]$ reappears in any of $L^d_{1..n'},\ $with$\ n'>n$ and label them accordingly.\footnote{This leaves unchecked profiles reserved on monitoring day 6. In practice, this does not affect our data analysis and results, see Sec.~\ref{subsubsec:res_data_diag}.}

\subsubsection{Sales prediction and listing reconstruction}
\label{subsubsec:meth_pred-and-simulation}

In this step, we use the resulting dataset to derive a sales prediction model as a function of the profile's features and employ it to simulate data for which we have no observations.

\smallskip
\noindent\textit{Modelling profile sales.}
To build our sales model, an important consideration is that attacker decisions to purchase a profile may be affected by what alternatives are available for selection at the moment the decision is taken~\cite{anderson1966consumer}.
As we cannot fully reconstruct this (ref. \LIM{4,5}), we model it at the level of the observation day $d$ as a random effect (see~\cite[Ch.13, pp. 489, for a formal definition, and 13.2.3 pp. 495 for a discussion on coeff. interpretation for cluster-specific models]{agresti2003categorical}) that captures the (time-dependent) stochasticity introduced, on the customers' decision, by the alternative options available in that (those) day(s).
We note that each monitoring day accounted for a sale requires considering the (random) effects caused by the availability of not-yet-sold profiles for all the previous days, increasing the overall uncertainty to model. 
This creates a trade-off, as it implies that for every additional monitoring day $n\in[1..6]$ included in the sample we necessarily remove observation days (i.e., those without a complete monitoring up to day $n$), and therefore profiles, from $\bigcup_{\forall d \in D}L^d_{0..n}$ (as the chance that at least one data collection failed increases with $n$, due to \LIM{1}). 
We therefore prioritize keeping modelling complexity at a minimum while retaining the highest number of data points for our model.\footnote{Due to the inherent uncertainty of the purchase decision process, we prioritize minimizing the True Negative Rate (TNR) of our estimator, and consider two different threshold values for sale prediction corresponding to $TNR = 95\%$ for the `conservative' estimator, and $TNR = 80\%$ for the `generous' one.} 

\smallskip
\noindent\textit{Data reconstruction and simulation.} 
For each day $d$ for which we have a data collection $L^d_0$ but no subsequent observation in $L^{d}_{1..6}$, we use the estimated model to predict which profiles appeared in that day were likely to be sold.
For every missing $L^d_0$, we (a) first estimate the number of products we should have collected for that day, and (b) run a simulation batch reconstructing which profiles could have appeared on that day. To have an estimate for (a), we consider the first available $L^d\textsubscript{1..6}$ to make a lower bound figure of how many profiles appeared in $L^d_0$, and derive our estimation by scaling it up by the average rate of sale at that monitoring day; if this information is not available as well 
we use the overall market recap (provided daily by \market) with the number of appeared profiles for that day \textit{d}. However, we find that this information is not always accurate as it reports fewer profiles than what we measure in $\approx 30\%$ of cases. Thus, we correct this figure by computing the average ratio between the measured offer and the numbers reported in the market recap. To perform (b) we build a set of simulations by sampling, with replacement, the number determined by (a) of profiles from the surrounding days.\footnote{This decision was taken after checking that profiles appearing on subsequent days have similar characteristics. We report results in the Appendix.} On the simulated data we then apply the estimated model to predict sales and calculate central estimates and confidence intervals of market statistics and sale trends from the resulting data distributions.
Due to computational constraints, we build two batches of simulations: one ($n=100$) retaining detailed data on sampled profiles (e.g., geographic location, available resources, ..), is used in Subsec.~\ref{subsubsec:trends-supply} and \ref{subsubsec:price-sensitivity} to report on detailed profile descriptors. For the second batch of simulations ($n=10'000$) we only retain chosen statistics from each simulation and use it to estimate the overall market value in Subsec.~\ref{subsubsec:market-value}. The high number of simulations here is chosen to provide as accurate an overall figure as possible of the sales data. In either case, simulated days are always clearly marked in the reported figures.

\subsection{Ethical considerations}

The details of available profiles advertised on \market\ before purchase do not contain any PII. Advertised profiles include a censored IP address (e.g., 14.25.xxx.xxx), country of origin, affected OS, and a list of the websites for which stolen credentials exist, with no details on said credentials. 
Similarly, available cookies are reported as a count per browser, alongside a list of the affected browsers. Because this study relies solely on information available in profile listings on \market, the collected data does not contain any PII. An ethical revision of this research was performed by the relevant board at our institution and approved under reference no. ERB2021MCS1.

\section{Results}
\label{sec:results}

We first describe the data preprocessing and the resulting overview of the market data; the section then continues by analysing attacker profile acquisition trends, estimating sale volumes, and market size. 

\subsection{Data preprocessing}
\label{subsec:res_dataprep}
\vspace{-1mm}
\subsubsection{Data collection and enrichment} 
\begin{figure}[t]
    \centering
    \includegraphics[width=0.81\columnwidth,keepaspectratio]{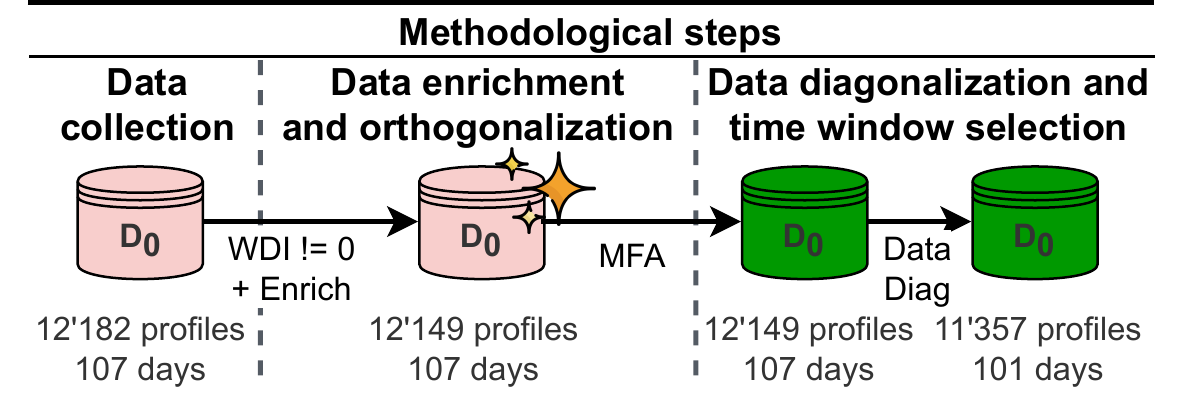}
    \vspace{-4mm}
    \caption{Data preprocessing pipeline.}
    \label{fig:pipeline}
    \vspace{-4mm}
\end{figure}

\vspace{-1mm}
Fig.~\ref{fig:pipeline}
provides an overview of the data preprocessing pipeline.
The data collection spans from Jan 21\textsuperscript{st} 2021 to Jun 30\textsuperscript{th} 2021\footnote{Crawling started in Nov 2020, but as \market\ went offline for an infrastructural upgrade from 11 Dec 2020 to 15 Jan 2021 we discard data from the previous period for consistency. Crawling resumed on the 21\textsuperscript{st} Jan.} 
and counts a total of $107$ complete $L^d_0$ over an observation period of a $161$ days, corresponding to a total of $12'182$ profiles. 
From the country of origin of each user profile, we derive the 2020 per capita GDP (Worldbank NY.GDP.PCAP.CD) indicated as WDI\footnote{If data from 2020 is not available for a country, we use the most recent estimation present in the same database.}. We found $33$ profiles originating from Reunion, Mayotte, French Guiana, Guadeloupe, and Taiwan, for which no information is available; we discarded them from the analysis, reducing the number of profiles to $12'149$.
Further, for each profile, we count the number of compromised browsers by family (e.g., Firefox, Opera), the available cookies by browser family, and the available resources (and related webplatforms) divided into six categories: \services, \commerce, \money, \other, \social\ and \crypto. Categories represent the purpose of the platform examined (e.g., \money\ contains websites of financial institutions enabling money transactions, \commerce\ includes platforms for e-commerce, ...). For consistency and benchmark, we adopted the same categorization scheme reported from \cite{campobasso2020impersonation}. We identified a total of $1'839$ distinct platforms. $576$ identifiers represent the same website or respective Android app (e.g., WellsFargo and \texttt{android://com.wf.wellsfargomobile/}); to avoid data duplication, we collapse those under the same identifier, reducing the number of distinct platforms to $1'297$. 
We assign each platform to its corresponding category. This yields $475$ platforms of type \services, $357$ \commerce, $265$ \money, $127$ \other, $39$ \social, and $34$ of type \crypto.
For each profile, we derive the number of resources in each category. 
Following the validation process outlined in Sec.~\ref{subsubsec:data-enrichment}, the final classification agreement was $97\%$. 
Tab.~\ref{tab:descstats} reports the dimensions of the resulting data. 

\vspace{-1mm}
\subsubsection{Feature extraction and orthogonalization}
\label{subsubsec:feat-extr}


The MFA analysis comprises overall $18$ variables (ref. Tab.~\ref{tab:descstats}). 
Variables are assigned to the groups \texttt{Price}, \texttt{Browsers}, \texttt{OS}, \texttt{WDI}, \texttt{Credentials}; \texttt{Sold} is considered only as a contrast variable and is not included in the MFA (as it represents an outcome and not a feature of the profile).
We log-transform and scale every numeric variable to unit variance, to ensure each variable equally contributes to the definition of the factor space. As for our application, the main purpose of the MFA is to get rid of multicollinearity issues across variables (as opposed to dimensionality reduction), so we do not constrain the number of dimensions in output of the MFA. We employ the \texttt{FactoMineR} package's~\cite{factominer}  \texttt{MFA} implementation in the statistical software package R. 
We run the MFA analysis on all the $12'149$ enriched profiles. We obtain $20$ orthogonal dimensions; for brevity, we report here the first $9$, representing $89.25\%$ of the overall variance (a full breakdown is available in the Appendix). 
Fig.~\ref{fig:mfa_contrib_plot}
\begin{figure}[t]
    \centering
    \includegraphics[width=0.86\columnwidth,keepaspectratio]{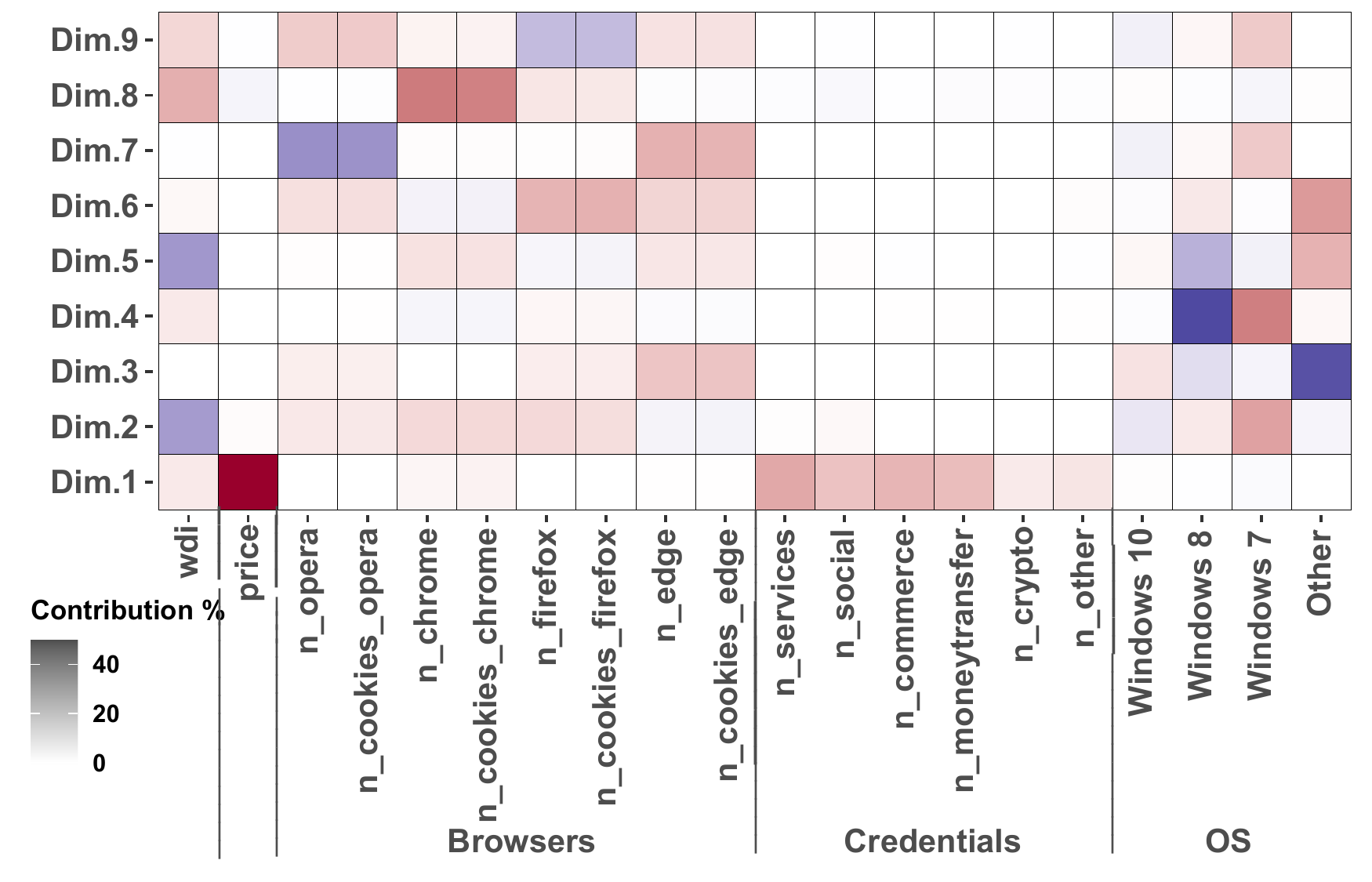}
    \begin{minipage}{0.9\columnwidth}\footnotesize 
    Blue and red respectively indicate positive and negative contributions of each variable to a dimension. Color intensity is proportional to the magnitude of the contribution.
    \end{minipage}
    \vspace{-1mm}
    \caption{Variables' contributions top 9 MFA dimensions.}
    \label{fig:mfa_contrib_plot}
    \vspace{-5mm}
\end{figure}
offers a full breakdown of the contributions from each variable for the resulting top $9$ dimensions. Each dimension is calculated as a linear combination of all variables; the coefficients assigned to each variable within a dimension (i.e., their `loadings') are correlated to each variable's contribution to that dimension. The sign of that coefficient indicates whether the variable and the dimension are positively (red) or negatively (blue) correlated. 
Fig.~\ref{fig:mfa_quanti_var} (reported in the Appendix, together with an extended description of MFA interpretation) provides insight into the construction of the MFA dimensions and the contributions of each variable within their groups. 
 To illustrate, we discuss the top $3$ addressing the most variance in the data. A closer look at Fig.~\ref{fig:mfa_contrib_plot} shows that the three variables within the \texttt{Creds} group \texttt{n\_moneytransfer}, \texttt{n\_services} and \texttt{n\_commerce} contribute the most, together with \texttt{Price}, to \texttt{Dim.1}. Therefore, \texttt{Dim.1} can be interpreted as representing high-resource, high-cost profiles within \market. That is to say, profiles similar in composition to the feature values captured by \texttt{Dim.1} (e.g., a high price) will score high on this dimension. 
Similarly, \texttt{Dim.2} is mostly influenced by profiles characterized from variables in the groups \texttt{Browsers}, \texttt{OS}, and \texttt{WDI}. \texttt{Dim.2} captures profiles from relatively poor countries according to the WDI index but rich in cookies. Interestingly, \texttt{Dim.2} also reveals that those profiles are more likely to exhibit older operating systems (Windows 8 and 7) and to feature browsers different from Edge. 
Profiles characterized by Edge running on Windows 10 instead seem largely captured by \texttt{Dim.3}.
Similar considerations on the profiles' characteristics can be made by comparing the interaction patterns visible in Fig.~\ref{fig:mfa_quanti_var} across all dimensions and variables (groups).

\vspace{-3mm}
\paragraph{Data diagonalization and time window selection.}
\label{subsubsec:res_data_diag}
The diagonalization process offers insight into the available data that can be used to model customer purchases. We evaluate the fraction of data that remains available to our modelling when varying the size of the measurement window for $L^d\textsubscript{0..n},\ n\in[1..6]$. Results are reported in Tab.~\ref{tab:available-data} in the Appendix, together with additional details on its construction.  
Among sold profiles, more than half ($58\%$) is sold within the first day. By contrast, the fraction of overall sales that can be accounted for by including subsequent days does not surpass 78\% of sales overall (including up to $L^d_{3}$), but at the price of removing 19 observation days (as opposed to 6 with $L^d_1$) from the sample and $\approx2'000$ profiles. 
These missing observations not only remove data for the model training but also create `holes' in the data collection that will have to be `filled back in' via model prediction, bringing in additional uncertainty.
To identify whether profiles of specific types are more likely to be sold after a certain number of days since their listing on \market, we look (not reported here for brevity) at the features of sold and unsold profiles. A set of Wilcoxon Sign-ranked tests finds no overlap across observations, suggesting that looking at profiles sold on a given day is representative of looking at those sold on surrounding days.

For these reasons, we consider only looking at the first day of sales as an acceptable trade-off. This results in the final dataset comprising $11'357$ profiles (of the $12'128$ originally fetched), sampled across $101$ (out of $107$) observation days while capturing 58\% of the overall sales.\footnote{This also excludes the data censoring our data diagonalization suffers from for profiles `reserved' on the $6\textsuperscript{th}$ monitoring day, discussed in Sec.~\ref{subsubsec:meth_data_diag}.} This gives us a total of six $L^d_0$ days with missing $L^d_1$ and $161-107=54$ missing $L^d_0$ days to simulate, for which we predict sale outcomes as detailed in Sec.~\ref{subsubsec:meth_pred-and-simulation}.



\vspace{-3mm}
\subsubsection{Overview of \market\ profiles}
\label{subsubsec:victims-overview}
\vspace{-2mm}
Tab.~\ref{tab:descstats} provides descriptive statistics of the final dataset.
\begin{table}[t]
\vspace{-3mm}
\centering
\caption{Descriptive stats for $L^d\textsubscript{0,1}$ and related MFA groups.}
\label{tab:descstats}
\scalebox{0.69}{
\small
\begin{tabular}{p{0.02\columnwidth}p{0.055\columnwidth}p{0.186\columnwidth}rrrr}
\toprule
& \textbf{Grp}  & \textbf{Variable}  & \textbf{Min} & \textbf{Mean} & \textbf{Max} & \textbf{SD} \\                             
  \midrule
  \multirow{13}{*}{\rotatebox{90}{\textbf{Original Variables}}}& Price & Price (USD) & 1 & 21.32 & 350 & 24.91 \\
  \cmidrule{2-7}
  & \hspace{2mm}\multirow{8}{*}{\rotatebox{90}{Browsers}} & \# Opera & 0 & 0.20 & 1 & 0.40 \\ 
  & & \hspace{3mm}\# cookies & 0 & 122.86 & 7332 & 501.58 \\ 
  & & \# Chrome & 0 & 0.76 & 1 & 0.43 \\ 
  & & \hspace{3mm}\# cookies & 0 & 1165.81 & 9448 & 1215.45 \\ 
  & & \# Firefox & 0 & 0.26 & 1 & 0.44 \\ 
  & & \hspace{3mm}\# cookies & 0 & 185.60 & 5911 & 601.31 \\ 
  & & \# Edge & 0 & 0.10 & 1 & 0.30 \\ 
  & & \hspace{3mm}\# cookies & 0 & 43.22 & 4098 & 253.67 \\ 
  \cmidrule{2-7}
  & OS & OS & -- & -- & -- & -- \\
  \cmidrule{2-7}
  & \multirow{3}{*}{--  $^\ddag$} & Date infect & 21-01-21 & 15-04-21 & 30-06-21 & 51.44 \\ 
  & & Date update & 21-01-21 & 15-04-21 & 30-06-21 & 51.45 \\ 
  & & Country & -- & -- & -- & -- \\
  \midrule
  \multirow{9}{*}{\rotatebox{90}{\textbf{Data Enrichment}}} & WDI & WDI & 126.90 & 26999.64 & 86601.56 & 18801.68 \\ 
  \cmidrule{2-7}
  & \hspace{2mm}\multirow{6}{*}{\rotatebox{90}{Credentials}} & \# Services & 0 & 10.78 & 569 & 16.56 \\ 
  & & \# Social & 0 & 4.09 & 263 & 7.36 \\ 
  & & \# Commerce & 0 & 3.17 & 149 & 7.11 \\ 
  & & \# MonTrnsfr & 0 & 1.38 & 248 & 4.96 \\ 
  & & \# Crypto & 0 & 0.18 & 53 & 1.21 \\ 
  & & \# Other & 0 & 0.25 & 38 & 1.08 \\ 
  \cmidrule{2-7}
  & Sold$^\dagger$ & Sold & -- & -- & -- & -- \\
\bottomrule
\multicolumn{7}{l}{\small{$^\dagger$ Supplementary variable of the MFA; $^\ddag$ Not part of the MFA.}}
\vspace{-8mm}
\end{tabular}}
\end{table}
Profiles are offered at an average price of $21.32$ USD; the $5\%$ most expensive profiles are priced at $59$ USD or more. When looking at sold profiles, the average price reaches $25.96$ USD, with the $5\%$ most expensive exceeding $101$ USD. Chrome appears to be the most popular browser among the affected victims, being on average 3 times more frequent than Firefox and Opera; Safari and Internet Explorer never appeared during the analyzed period. On average, profiles contain predominantly \services\ credentials, followed by \social\ and \commerce. Since the data collection happens at most $24$ hours after a profile has been published, the date of the last update for each profile often matches the infection date; the former tells a customer whether a profile contains fresh credentials, and it is relevant when looking at older profiles. 

By looking at the reported standard deviations, there are large variations in the number of stolen cookies across all browsers. This difference suggests that the target population shows diverse traits in terms of Internet usage: the $90\%$ of the victims found in $L^d\textsubscript{0,1}$ appear to use few services and few platforms only, counting $48$ credentials or less, while the remaining $10\%$ has $87$ credentials on average and $883$ at maximum. Similar considerations on the number of stolen credentials may shed light on some population characteristics. While a small number of credentials per profile may indicate a limited Internet activity of the victim, when paired with profiles presenting a large number of cookies it may indicate users not saving their passwords in the browser, or using a password manager.\footnote{Widespread infostealer malware like AZORult and RedLine are incapable of stealing passwords from password managers, although an attacker could identify the master password by sniffing keystrokes\cite{kaspersky_2021,andrioaie_2022}.} Looking at the geographical distribution of profiles, the overwhelming majority of profiles originates from Europe ($62.14\%$), followed by North America ($11.97\%$), South America ($11.83\%$), and Asia ($11.01\%$)\footnote{As it is often the case with Russian-based cybercriminal ventures, profiles in our collection do not concern countries in the Russian area of influence (Commonwealth of Independent States), except for Kyrgyzstan and Uzbekistan. A deeper look into the market shows that among the ten CIS countries, only Belarus, Kazakhstan, and Russia do not appear at all.}; Africa and Oceania together account for the $3.04\%$ of total profiles. Profile composition varies across regions; North American profiles are generally richer in credentials. These profiles offer, on average, $27$ credentials, while Europe, South America, and Asia offer respectively $20$, $19$ and $13$. The same trend is noticeable also with \commerce\ and, albeit less remarkably, with \crypto\ credentials. That is well reflected in the price of these profiles (respectively, on average $34.22$, $20.29$, $19.91$, and $14.03$ USD), following the intuition that wealthier countries have a more appealing resource composition for the market's customers, confirming the findings of~\cite{campobasso2020impersonation}.

\subsection{Attacker activity on \market}
\label{subsec:res_sales}

In this section, we provide an analysis of attackers' purchasing decisions and associated factors. Unless otherwise stated, reported significance statistics are produced via a batch of Wilcoxon Rank-Sum tests; we consider an $\alpha$ value of $5\%$ as the threshold for statistical significance.

\vspace{-3mm}
\subsubsection{Analysis of attacker preferences}
\label{subsec:disc_model_sales}

To evaluate customer preferences when selecting profiles to buy among those offered on \market, we define a set of nested generalized linear mixed models (GLMM) to estimate the relation between the obtained profile dimensions and a purchase decision. 
We build the final model including dimensions in output of the MFA in incremental steps, ordered by their relative contribution in explaining our dependent variable (i.e., sales; details on this process in the Appendix). 
The final model obtains an $R^2$ of $27.8\%$. The model construction assures that virtually all the information available in the market data is captured.\footnote{To further enrich the model, one would have to look beyond the market data itself and, for example, interview the customers of the market when they make a purchase decision. That is an inherent limitation common to all studies of this type (see, for example, discussion in \cite{soska2015measuring,aliapoulios2021swiped}); the decision of providing conservative sale estimates derives from this observation, to avoid overshooting in presence of structural uncertainty in the data.}
The model obtains a satisfactory AUC of $0.77$, 
despite the high uncertainty inherent to the effect it models. For this discussion, the table below reports the dimensions that explain at least $1\%$ of the total variance in the model (\% of explained variance by each dimension reported below the coefficients).\footnote{
Because of the high uncertainty in the data, we employ all dimensions that an ANOVA test (not reported because of space constraints) evaluates as significant to be included in our sales prediction model. This is to maximize the model's power in predicting sales in our simulations for the missing data. However, for the purpose of interpreting results, we note that only dimensions that capture a large enough variance in the outcome are worth considering to add meaningful insights into attacker preferences. 
} Full details on the model are provided in the Appendix.
\scalebox{0.84}{
\centering
\begin{tabular}{p{0.11\columnwidth}p{0.09\columnwidth}p{0.11\columnwidth}p{0.09\columnwidth}p{0.09\columnwidth}p{0.09\columnwidth}p{0.08\columnwidth}p{0.13\columnwidth}}
\toprule
$c$ & \texttt{Dim.8} & \texttt{Dim.2} & \texttt{Dim.13} & \texttt{Dim.9} & \texttt{Dim.4} & \texttt{Dim.6} & \texttt{Dim.5}\\
\midrule
$-2.51^{***}$ & $0.62^{***}$ & $-0.41^{***}$ & $1.02^{***}$  & $0.32^{***}$   & $0.19^{***}$   & $0.38^{***}$     & $-0.17^{***}$ \\
--     & $(8.2\%)$      & $(5.7\%)$      & $(3.0\%)$       & $(2.7\%)$       & $(1.7\%)$       & $(1.6\%)$       & $(1.5\%)$ \\
\midrule
\multicolumn{8}{l}{\footnotesize \#obs = $11'357$, $R\textsuperscript{2}_m$ = $0.264$, $R\textsuperscript{2}_c$ = $0.278$, $std(c|day) = 0.25$, $^{***} p<0.001$} \\
\bottomrule
\end{tabular}}

\smallskip
\noindent Coefficients can be interpreted on the same scale; coefficients should be interpreted jointly with the dimension compositions reported in Fig.~\ref{fig:mfa_contrib_plot}. 
Positive (negative) regression coefficients mean that user profiles that score high on that dimension have a greater (lower) chance of being sold. The sign of the variable loadings for a given dimension (color-coded in Fig.~\ref{fig:mfa_contrib_plot}) indicates whether a variable is associated with a `high score' on that dimension depending on its value in the original distribution (i.e., above or below the mean).
For example, the positive coefficient of \texttt{Dim.8}, together with its variable compositions reported in Fig.~\ref{fig:mfa_quanti_var}, suggests that profiles with high \texttt{WDI} with a large number of cookies originating from Chrome, and Firefox to a lesser extent, are preferred by the attackers. 
By contrast, the negative coefficient for \texttt{Dim.2} suggests that profiles from less wealthy countries featuring older operating systems (Win 7,8) are less likely to be sold even if they might be high in resources/cookies.
\texttt{Dim.13} (dimension composition observable in Fig.~\ref{fig:mfa_contrib_plot_total} in the Appendix) suggests that attackers are interested in profiles rich in \texttt{Social} and \texttt{Moneytransfer} credentials, as long as they are cheap; 
\texttt{Dim.9}, \texttt{Dim.4} and \texttt{Dim.5} further corroborate that attackers prefer profiles originating from wealthier countries and characterize different profile configurations; \texttt{Dim.9} and \texttt{Dim.4} identify a group of profiles originating from systems running Win 7, but with a different resource composition from that of \texttt{Dim.2}. In particular, \texttt{Dim.9} suggests that Win 7 profiles are more likely to be sold if featuring data for Opera or Edge. Interestingly \texttt{Dim.5} and \texttt{Dim.6} indicate that the presence of a specified OS loses importance (`OS=Other') provided that the profile is associated with a high WDI and has several resources for Chrome or Firefox. 

Overall we find a positive association between the composition of profile characteristics and its likelihood of sale, with \texttt{WDI}, \texttt{Price}, and technical features such as the browser playing a predominant role in the purchase decision. Perhaps surprisingly, the type and number of included resources (e.g., \texttt{Social}, \texttt{Moneytransfer}, ..) seems to play only a limited role in the final decision. 
This may indicate that the average attacker does not necessarily prefer profiles rich in resources, as the victim's intrinsic value (i.e., their wealth, which may become available after a successful impersonation attack) is the same regardless of the type or number of associated resources (as long as the attacker has access to some of them). Indeed, the model coefficients indicate (across all dimensions, save for \texttt{Dim.13}) a strong attacker preference for profiles from high-WDI countries, suggesting that the perceived value of the victim's profiles is more relevant to the attacker than the number of ways in which that value can be accessed.  

\vspace{-2mm}
\subsubsection{Trends in market supply and attacker demand}
\label{subsubsec:trends-supply}

\begin{figure*}
    \centering
    \includegraphics[width=0.43\textwidth,keepaspectratio]{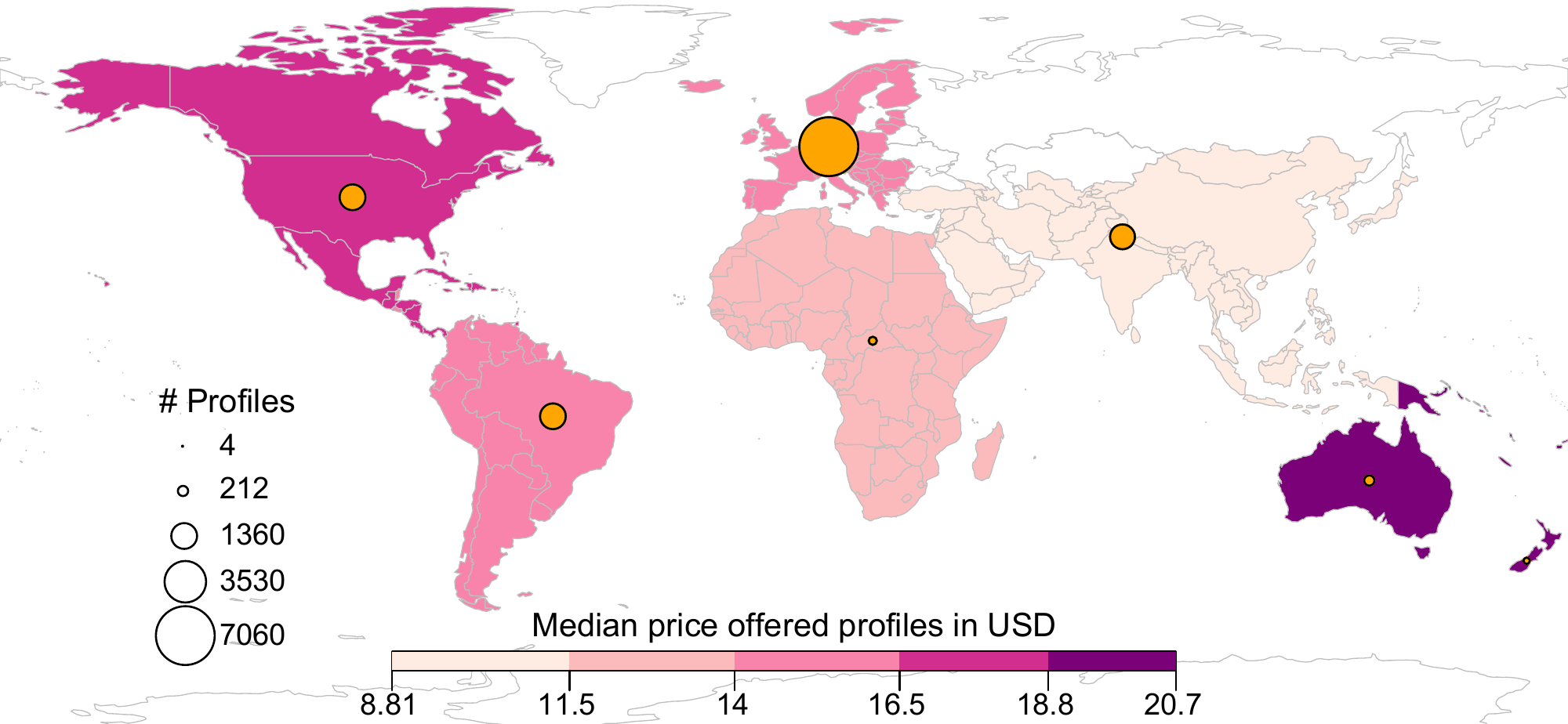}
    \includegraphics[width=0.43\textwidth,keepaspectratio]{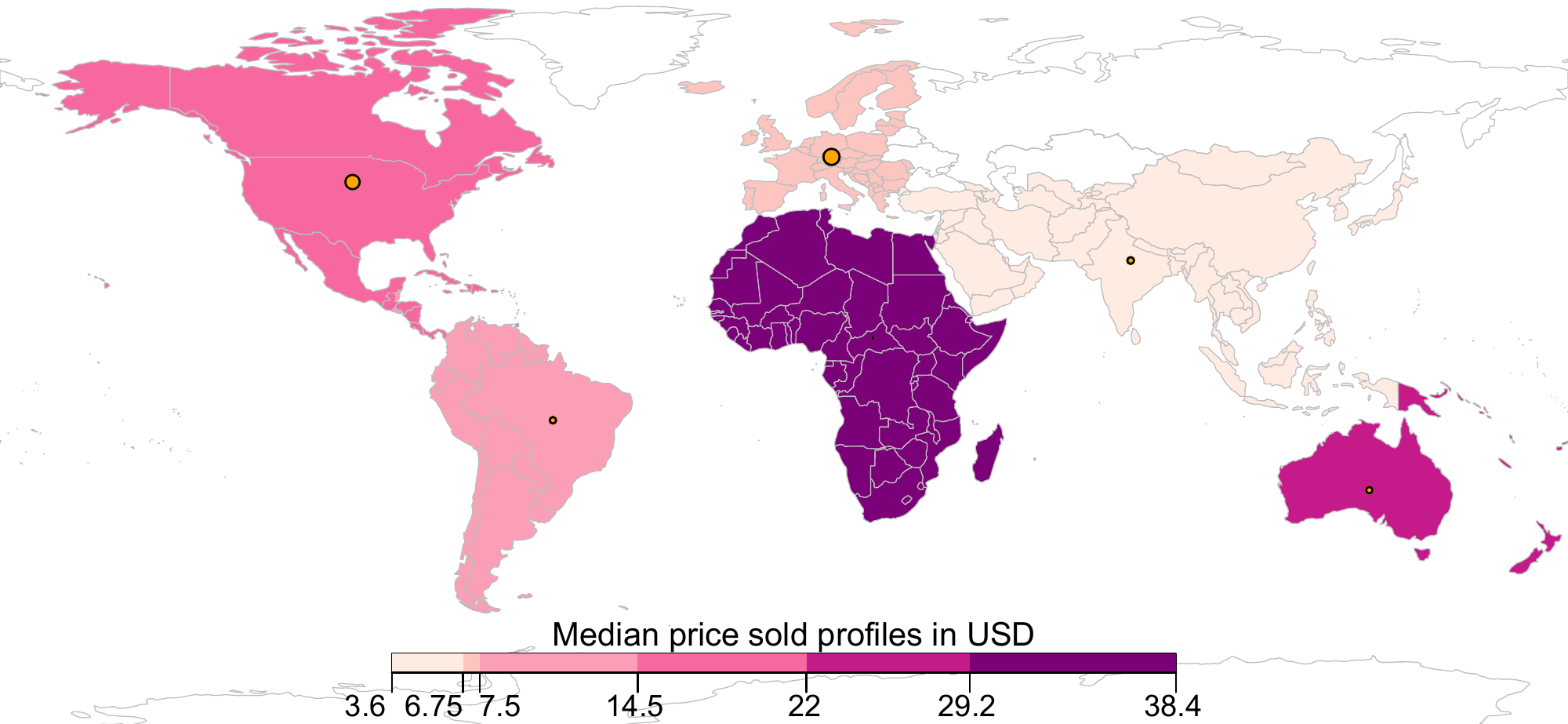}
    \caption{Overview of offered profiles (left) and acquired profiles (right) globally. Region colors represent median prices; superimposed dots represent volume of (either available or sold) profiles in the region. Comparing statistics from left to right provides an overview of profile offer and demand across regions.}
    \label{fig:world_map}
    \vspace{-2mm}
\end{figure*}

Fig.~\ref{fig:world_map} 
provides a bird's eye view of the volume and average prices for available and sold profiles, globally. 
The median price for offered profiles in Europe is $15$ USD, while in North America reaches $18$ USD, suggesting that profile composition is richer in the latter. The most expensive profiles originate from Oceania, with a median price of $19.5$ USD, although 
they are a minority ($0.93\%$). As per Subsec.~\ref{subsec:disc_model_sales}, and confirming results in 
\cite{campobasso2020impersonation}, profiles originating from wealthier countries show higher prices on average due to their per capita GDP (the only two countries attacked, Australia and New Zealand, have respectively 9\textsuperscript{th} and the 21\textsuperscript{th} highest per capita GDP in 2021). When looking at sold profiles, the demand sharply rises for North America, accounting for $34.54\%$ of total sales, while Europe `only' for $43.67\%$, suggesting that attackers' relative demand for North America's profiles is four times higher than that for Europe. This difference is well reflected in the median prices of sold profiles across the two regions; if the gap in median prices between North America and Europe is $18-15=3$ USD, when looking at sales this significantly widens to $21-7=14$ USD. That suggests a clear preference for attackers in North American profiles over European ones, even if supply in the latter is almost six times larger than for the former. Africa shows the highest profile median price ($35.5$ USD), but accounts only for $4$ sales in our sample.

\begin{figure}[t]
    \centering
    \includegraphics[width=0.9\columnwidth,keepaspectratio]{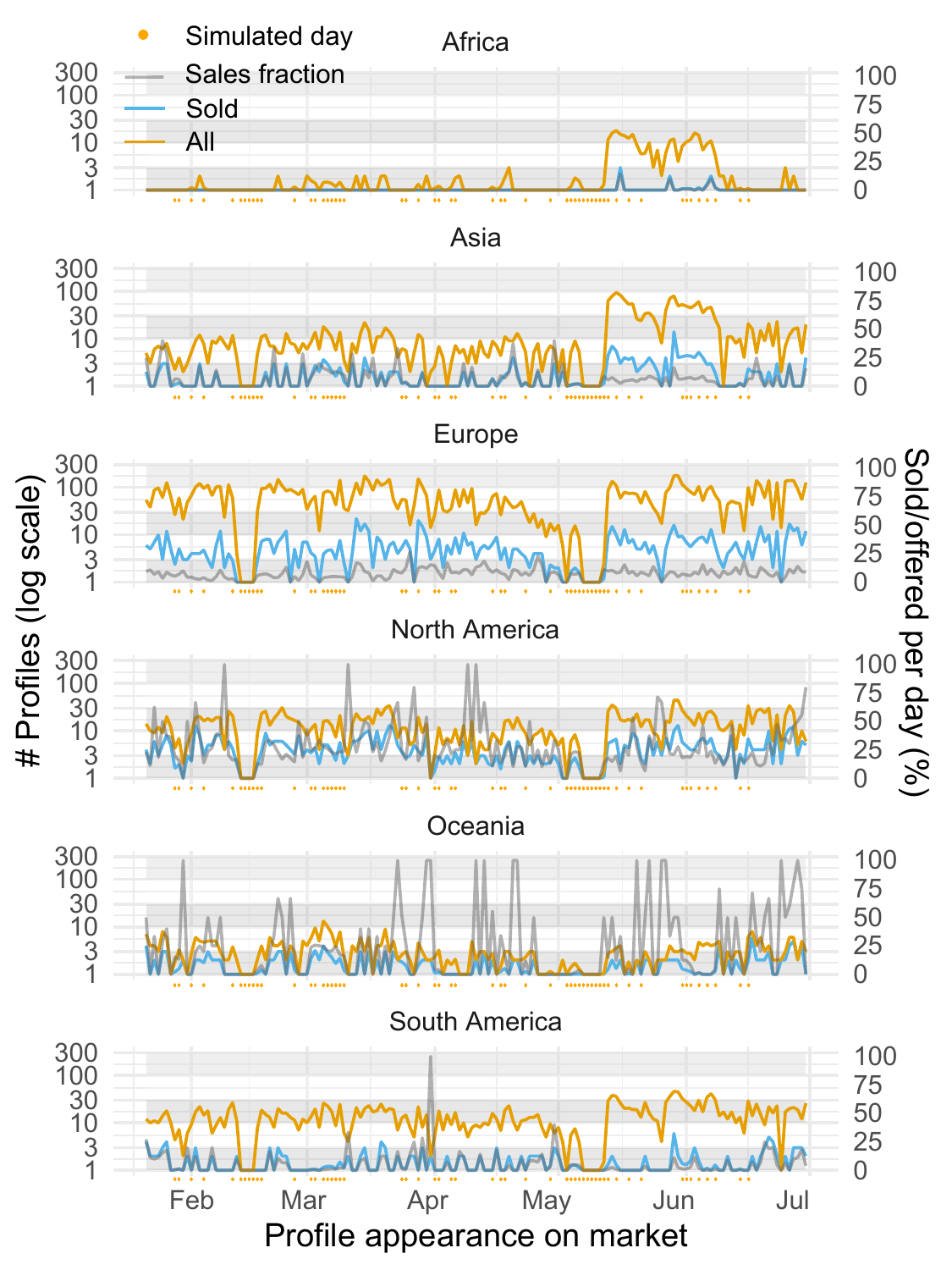}
    \vspace{-3mm}
    \caption{Timeline of (average) daily profile offering and sales by region.}
    \label{fig:profiles_vs_characteristcs}
    \vspace{-6mm}
\end{figure}

Fig.~\ref{fig:profiles_vs_characteristcs} 
provides an overview of rates of offered profiles (yellow line) and sold profiles (blue line) across regions. Market supply is not constant overall and shows highs in late February, mid-March, and between May and June. 
From mid-March to mid-April, North America's offer is scarce, with an almost matching demand. Interestingly, albeit Europe shows the same decrease in supply, demand remains stable, moving the fraction of sold profiles roughly from $8\%$ to $25\%$, suggesting that attackers could have bought some European profiles to make up for the shortage in North American ones. The same phenomenon is evident in Oceania, where the limited demand is often saturated in several days. In the second part of April, the trend partially reverts, with Europe's supply in strong decline (top early April $\approx100$ profiles/day, bottom late April $\approx15$ profiles/day, $p<0.0001$) and North America still declining but at a slower pace (top early March $\approx15$ profiles/day, bottom late March $\approx10$ profiles/day, $p<0.0001$). Overall, we observe a clear correlation between supply and demand for Europe (Pearson $cor=0.74,\ p<0.0001$) and North America (Pearson $cor=0.76,\ p<0.0001$); looking at the gap between the offered and sold curves, for North America we observe a higher fraction of sold profiles when compared to Europe and, in general, other regions. Among the latter, despite comparatively low volumes of provided profiles, Oceania appears to attract attackers. 
Looking at the fraction of sold profiles, it appears that Asia gathers similar interest compared to Europe, despite being roughly underrepresented by a factor of $10$ from the beginning of the observation period to the end of April.
Asia shows a significant increase in supply after the market shut down in early May ($p<0.0001$). Albeit South America provides similar amounts of profiles compared to North America, demand appears to be relatively low, with a few exceptions for some profiles in periods of particular shortage of profiles from other regions, such as mid-March to late April. Finally, Africa appears to be the most underrepresented region, with relative spikes in supply around mid-May and early June, leading to some of the only measured purchases we measured found in our observation period. No sale observation for Africa has been measured from February to the end of May 2021, although this may be partially explained as a byproduct of the adopted sampling mechanism whereby rare resources are likely not to be selected.

\vspace{-2mm}
\subsubsection{Attackers price sensitivity across profile types} 
\label{subsubsec:price-sensitivity}
\begin{figure*}[t]
    \centering
    \includegraphics[width=0.9\textwidth,keepaspectratio]{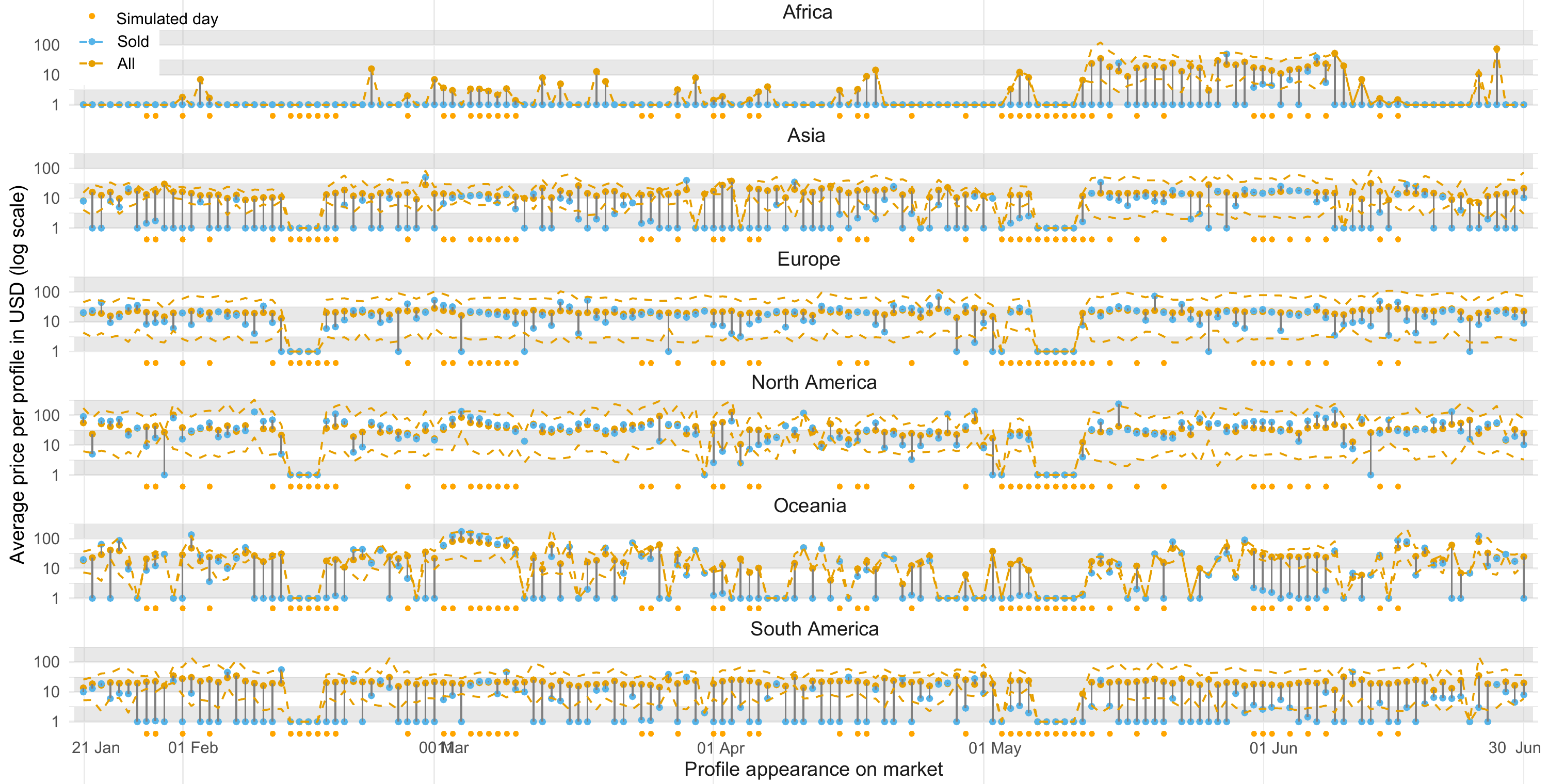}
    \caption{Average prices of offered (yellow dots) and sold (blue dots) profiles across geographical regions. Yellow dashed lines indicate 95\% confidence intervals of available profile prices.}
    \vspace{-3mm}
    \label{fig:timeline_price_region}
\end{figure*}

We now investigate how price-sensitive buyers are when choosing profiles with certain characteristics.
Fig.~\ref{fig:timeline_price_region}, 
reports average prices for provided and sold profiles across regions.
Supply and demand in North America exhibit modest prices in the offer compared to other regions from mid-May to the end of the observation period, while sales reach spikes of average price per sold profile as high as $300$ USD in mid-May. The average sold profile in North America is oftentimes as expensive as the top 5\% of provided profiles (blue dots in the region or above the dashed line indicating 95\%CI in the figure). 
By comparing trends in sales reported in Fig.~\ref{fig:profiles_vs_characteristcs} and prices in Fig.~\ref{fig:timeline_price_region}, it emerges that the North American profiles from late March to late April result in attackers purchasing almost the entirety of the daily supply. Throughout the observation period, the average price for North American sold profiles is higher than the corresponding offer, despite the baseline price being higher than in other regions ($sold=42.59,\ all=34.77$).
From February to April, European profiles gained some traction in sales, increasing the average sale price from $\approx10$ USD to $\approx30$ USD. 
By contrast, European profiles sold between April and early May are less on average ($m=3.52,\ sd=2.36$) than the previous period ($m=6.46,\ sd=5.08,\ p=0.004$), while supply prices remain rather stable (April to early May $m=19.20,\ sd=2.27$, late February to March $m=20.27,\ sd=1.76,\ p<0.001$).
After a general decline up until May in Europe and North America, profiles originating in Asia soared in volumes both in terms of the offer and demand up until early June, together with a renewed interest in North American profiles until the end of May.
After this period, sales volume in Asia started declining again, and North America and Europe became again the leading source for attractive profiles. 
Interestingly, profiles originating from South America present similar prices to Europe and offer the same volumes as North America, but they rarely seem to interest attackers, resulting in generally low sale prices.
That may reflect a general perception that profiles originating from that region are of low interest to attackers, who are willing to spend comparatively less to acquire those identities. 
When looking at Oceania, average prices for supply and demand move erratically, possibly due to the scarcity in the former; we witness a few notable sales reaching $100$ USD on average per day during late May and June. Finally, Africa shows significantly lower prices in the supply until the early May shrink. From mid-May, prices grow to Europe levels for roughly a month, but sales do not gain traction. 

\vspace{-3mm}
\subsubsection{Overall profile acquisitions and market value}
\label{subsubsec:market-value}
\vspace{-2mm} We now report overall sales trends and market revenues estimated from the described sales data. We report both `conservative' and `generous' sale estimations (model threshold at $TNR=95\%$ and $TNR=80\%$ respectively); 
the analysis reports between $1'799$ ($95\% CI = [1'757, 1'843]$) and $2'518$ ($CI = [2'462, 2'575]$) sold profiles out of $17'171$ ($10.5\%-14.7\%$ of offered profiles sold). 
Recall that we are collecting a random sample of the actual \market\ listings (ref. Sec.~\ref{subsubsec:meth_data_collect} and Sec.~\ref{app:data_reconstr_sim} in the Appendix for additional details), and measure only approximately half of the actual sales (ref. Sec.~\ref{subsubsec:res_data_diag}). To obtain a rough but realistic estimate of actual numbers, the reader can simply scale up reported figures by a factor of 10 (a more detailed review of this factor is provided in the Appendix). 
Fig.~\ref{fig:revenue_strict} 
\begin{figure*}
    \centering
    \includegraphics[width=0.85\textwidth,keepaspectratio]{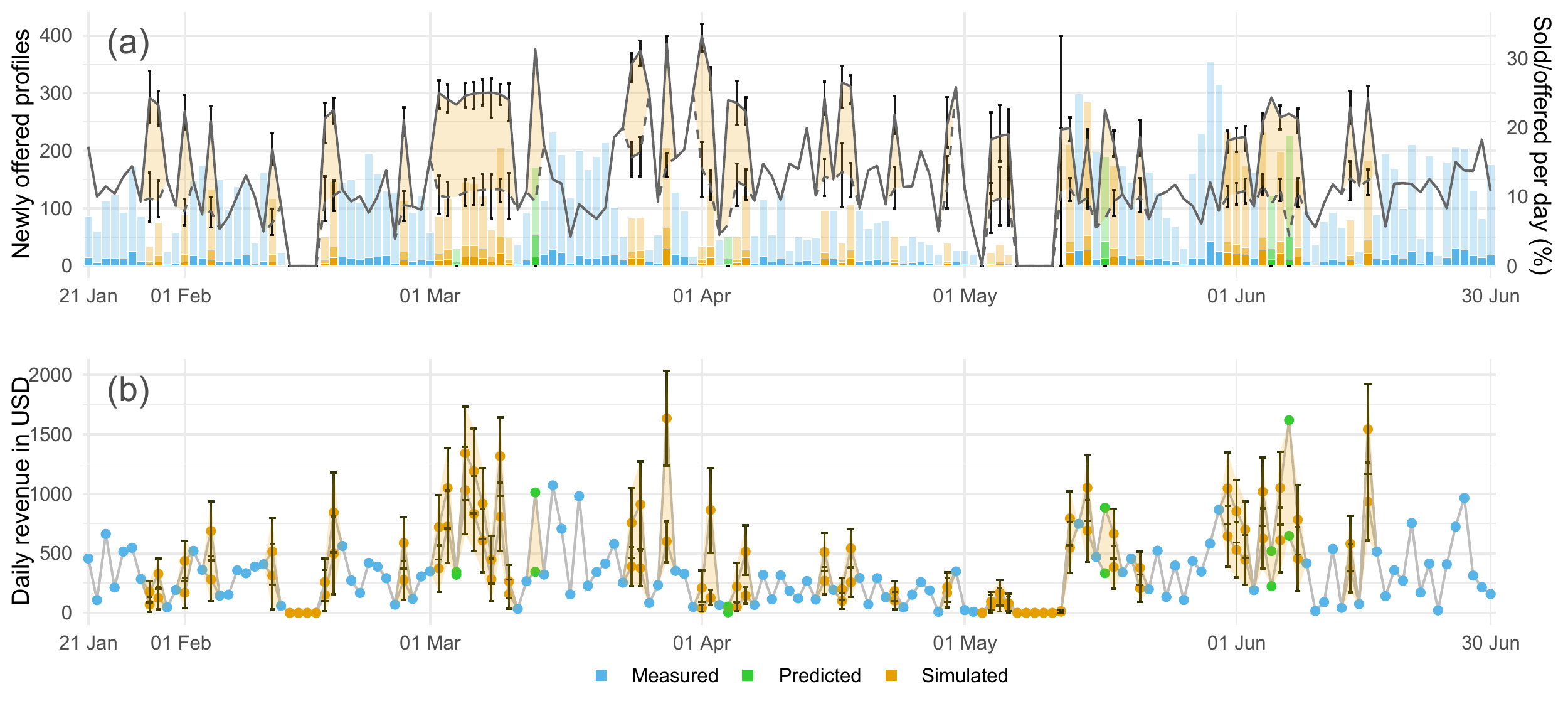}
    \begin{minipage}{0.87\textwidth}
    \footnotesize
     Estimates relative the market sample. To obtain a rough but realistic figure of actual sale revenues and volumes, one can scale reported quantities by a factor of 10. Shaded areas indicate the range between conservative estimates and `generous' estimates. Simulated sales (ratios) are reported with their respective standard deviations.
    \end{minipage}
    \vspace{-1.2mm}
    \caption{Measured, predicted and simulated sales volume (a) and daily revenues (b) on a sample of overall profiles.}
    \vspace{-3.1mm}
    \label{fig:revenue_strict}
\end{figure*}
reports a daily breakdown of the overall sales. Fig.~\ref{fig:revenue_strict}(a) provides an aggregate overview of newly available and sold profiles per day (respectively light-shaded for `generous' estimates and dark-shaded for conservative estimates for simulated/predicted days, and solid stacked bars for sold profiles) and the respective fraction of sold profiles (dashed line for conservative estimates, solid for `generous'); Fig.~\ref{fig:revenue_strict}(b) reports daily revenues, reporting values for estimates for both simulated and predicted days. Here we report numbers from the analysis next to the scaled figures in parentheses.
Looking at Fig.~\ref{fig:revenue_strict}(a),
overall supply sharply varies across periods, with profile provision ranging between 20(200) to a maximum of about 350(3500) in late May 2021, with sales peaking in the same period to 43(430) profiles per day. Periods of low/no supply are visible: next to market downtimes mid-February and early May, the profiles supply between April and May is very low overall, ranging from 120(1200) to less than 10(100) before completely terminating for 5 days. Interestingly, looking at daily sale patterns (trendline), the aggregate effects of sales during this period do not identify those effects of demand almost matching the offer as observed in the regional breakdown from Fig~\ref{fig:profiles_vs_characteristcs}, but rather it is true the opposite: the fraction of sold profiles amounts to $\approx25\%$ at the beginning of April and bottoms to $\approx12\%$ by the beginning of May; this suggests that attackers still seek profiles with peculiar characteristics and in case of scarcity they are not tempted from less appealing profiles, suggesting they are strategic in victim selection.
Some notable peaks are from mid-March to mid-April, and late May and June.  
Looking at daily revenues in Fig.~\ref{fig:revenue_strict}(b), we see an inflow averaging around $304$($3'040$) USD/day with peaks at approximately $720$($7'200$) USD/day from the conservative estimate, and $399$($3'990$) USD/day with peaks of $1'640$ ($16'400$) USD/day for the generous one.
Daily revenues largely reflect sale volumes and appear to cycle between high-demand periods (March and June 2021), and lower-demand ones (Jan/Feb and April 2021). 
\vspace{-4mm}
\paragraph{Estimate of market size and revenue.}
From our data reconstruction, we estimate (conservatively) that,
over the period of $161$ days from Jan 21\textsuperscript{st} 2021 (incl.) and Jun 30\textsuperscript{th} 2021,
\market\ published overall $\approx 97'655$ advertised profiles, $\approx 20'000$ of which sold within the first day ($95\%\ CI = [19'572, 20'530]$) at an average 
price of $\approx27$ USD ($[25.83, 28.64]$). Overall, we estimate that the total revenue for \market\ during the reported period is of $\approx 540'000$ USD ($[517'729, 574'118]$) (i.e., about $1.2m$ USD/yr, assuming that the observation period is representative of the unobserved one). A less conservative estimate ($TNR=80\%$) results in $\approx 28'000$ profiles sold ($[27'426, 28'685]$) at an average price of $25.48$ USD ($[24.25, 26.77]$), for a total revenue in the observation period of $\approx 715'000$ USD ($[680'312, 750'884]$).



\vspace{-3mm}
\section{Discussion and conclusions}
\label{sec:discussion}
\vspace{-2mm}
\begin{figure}[t]
    \centering
    \includegraphics[width=0.92\columnwidth,keepaspectratio]{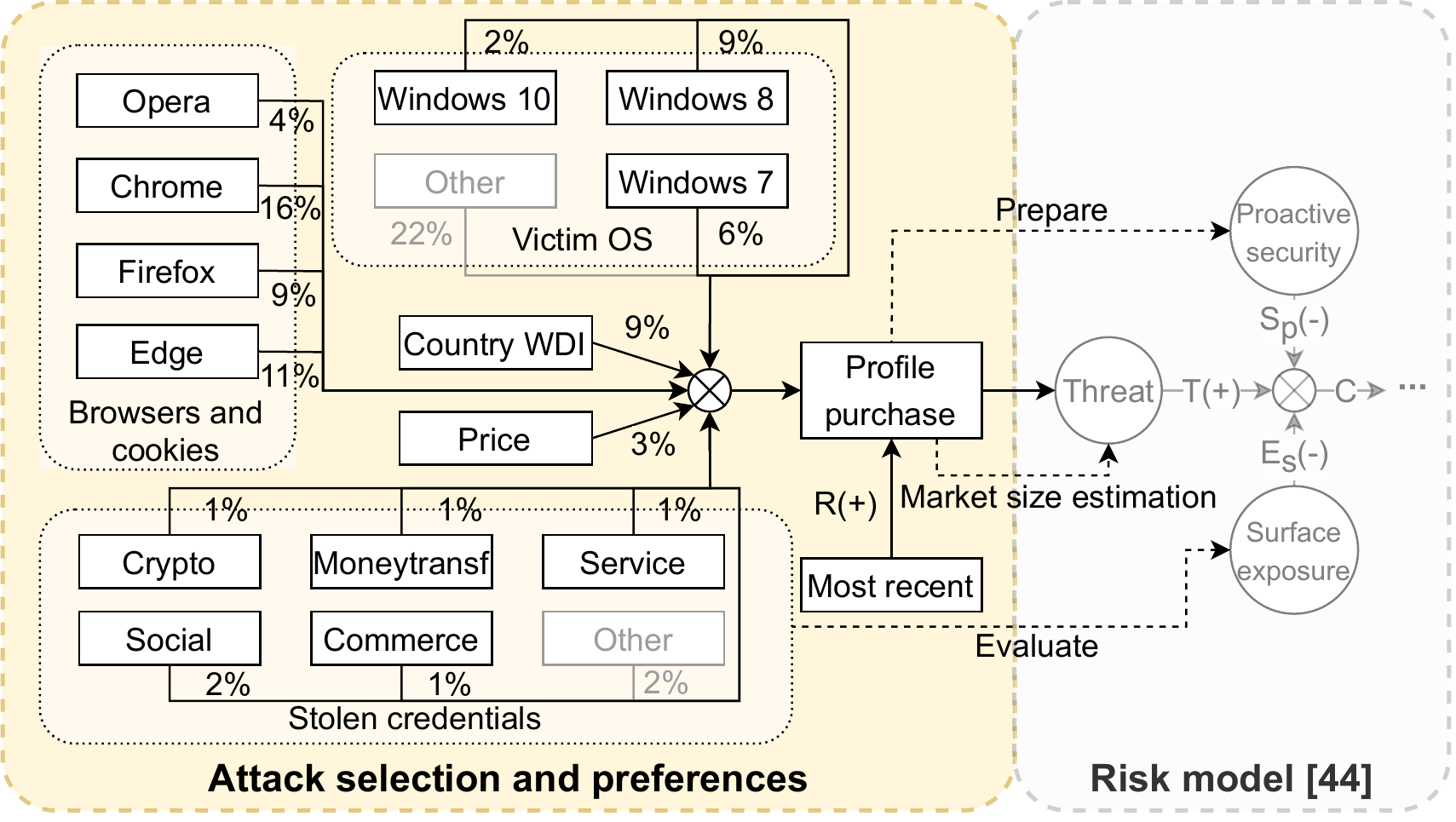}
    \vspace{-2.5mm}
    \caption{Attacker preferences within \impaas.}
    \label{fig:theoretical_framework}
    \vspace{-4mm}
\end{figure}
\textit{Attack selection and preferences.}
The first observation emerging from our analysis is that attacker decisions and preferences within the \impaas\ threat model are complex: effects cannot be synthesized and quantified at the level of single factors.
Rather, the attacker decision can be better modeled by accounting for the interactions across different profile characteristics. For example, we find that a profile low on resources may still be attractive if running on a recent OS, and belonging to a profile from a wealthy country. This suggests that \impaas\ attackers may prefer high chances of success over having a wide attack surface (e.g., potentially targeting many online resources/websites within a profile). Because of this complexity, one cannot quantify and isolate the effect of a rise of one point in a variable alone (e.g., `WDI') on the odds of purchase (differently, our model can be used directly to evaluate what is the probability of purchase of a specific profile configuration). However, by analysing the relative contribution of each factor across dimensions, and the importance of those dimensions in explaining the observed outcome (i.e., a sale, in terms of the change in $R^2$ for which that dimension is responsible), we can still derive a first indication of the \textit{relative importance of each factor in the final decision}. \footnote{This is different from assigning a given variable a signed coefficient quantifying its effect on odds of sales. Rather, this quantifies the variance in the final decision captured by a specific variable, across dimensions.}

Fig.~\ref{fig:theoretical_framework} reports the results within the overall framework of Woods and Böhme's risk model\cite{woods2021sok}. The figure offers a breakdown of the original variables involved in the attacker decision process, and reports the variance on sales they explain across all dimensions.
From the analysis,\footnote{For completeness, in the figure we also report the categories `Other' for both the \texttt{OS} and \texttt{Credentials} groups. However, as `other' is a bin variable for which no clear classification emerges, we refrain from making conclusions. The high relevance of OS=other (i.e., OS is \textit{not} specified) is due to  almost all of the associated 122 profiles being sold. That suggests that this is an artifact of the data rather than a specific effect worth capturing.} it emerges that the wealth of the country from which the profile originates is an important factor attackers consider when making a purchase decision (capturing $\approx 9\%$ of its total variance). By contrast, the price of a profile only plays a minor role in the decision (3\%), perhaps as a result of the profiles being overall relatively inexpensive. Interestingly, purchase decisions seem to be highly affected by the browser from which the stolen information and cookies originate. Google Chrome accounts by itself for 16\% of the variance in the purchase decision, followed by Edge and Firefox (at approximately 10\% each). Opera seems to be the least relevant browser in the decision. The high relevance of Chrome in the purchase decision may be confounded by \market\ providing their browser extension for Google Chrome itself (ref. Sec.~\ref{subsec:related_work}), perhaps increasing an attacker's confidence that the purchased profile will work on their setup. Overall, the type of browser and the cookies they come with account for approximately 40\% of the overall variance.
The OS also plays an important role ($\approx 17\%$), possibly indicating a selection mechanism that disregards older systems, as seems to be consistently (i.e., across all dimensions) suggested by the sales prediction model (Subsec.~\ref{subsec:disc_model_sales}).
Surprisingly, the composition in credentials accounts for a minority of the total variance ($\approx 6\%$), suggesting that these play a relatively minor role in the final decision. An explanation for this may be that most profiles are `rich enough' in resources of different types, meaning that relative differences across profiles do not impact much the final decision. That is also in line with the notion of rational `mass attackers' looking for any target for which their attacks will work, as opposed to specific targets~\cite{allodi2022work}, particularly when facing high costs to monetize the attack~\cite{herley2012nigerian}.

Perhaps unsurprisingly, we can also conclude that a key factor in the purchase decision is how recent the information within a profile is. An explanation is that attackers may believe that information within more recent profiles (e.g., a token within a cookie) is more likely to still be valid at purchase time. This however emerges only informally from the initial data analysis, as opposed to formally from the sales model (which only accounts for profiles sold on the first day).

\smallskip
\noindent\textit{Proactive security and surface exposure.} Understanding attacker preferences provides awareness on the possible risks connected to \impaas. 
For example, an organization could monitor \market, or any other emergent \impaas\ service or provider, to gauge the level of exposure of their employees (e.g., through the presence of an employee-only login portal website amongst available resources) to possible attacks. 
We note that to do this, the organization needs not buy specific profiles: \market\ provides the list of the (sub)domains for which credentials are available as part of the profile description. As sold accounts tend to be traded within a day, any preventative action should be taken swiftly, and can be enforced only temporarily to minimize negative externalities on final users. For example, when observing the appearance of profiles for that organization (and/or predicting their sale), risk-based authentication mechanisms could be temporarily disabled or hardened to require second factor authentication in all cases for the upcoming period. Similarly, observing or predicting a sale for a profile with credentials for that organization may be communicated to central monitoring services (e.g. a Security Operation Center monitoring the infrastructure) to raise alert levels around suspicious login actions.
Further, the geographical information of a profile could inform different branches, for example to prioritize internal audits looking for affected employees. Further research may look at how to integrate `live' IoCs from underground markets in security processes. For example, sale predictions could be further used to prioritize specific responses, or evaluate risk levels.
Finally, an organization could consider investigating the risks posed to their specific RBA configuration. This may be achieved by acquiring profiles featuring compromised corporate (employee) accounts (barring any required legal checks), and identifying the corresponding infected devices in the organization. 

\smallskip
\noindent \textit{Market size estimation.} Findings related to the size of underground markets are oftentimes a precursor to law enforcement initiatives such as takedown actions. Evaluating the number of sales of a market is a rare opportunity requiring either market infiltration or usually only coming after the market has already suffered from some shock (e.g., a leak or hack). Our investigation reveals that in approximately six months, \market\ made available data on $\approx100k$ Internet users; $20k$ have been sold, and therefore likely attacked, by \market\ customers in the same period. The sales activity indicates a remunerative business model, especially considering \market\ is a single-vendor market (as opposed to a market platform~\cite{soska2015measuring}).

\smallskip
\noindent \textit{Lesson learned in measuring underground activities.}
Whereas our data collection methodology is tailored to \market\, it may also inform the design of other measurement methods addressing these or similar challenges. In particular, the community could attempt to address these challenges systematically to produce robust and reusable software for stealth underground monitoring, helping other researchers to tap data from the underground. In retrospect, features that could have eased the data collection process include a system to manage the unreachability of the market; among these, attempting to `greedily' (i.e., as soon as possible) collect data could be viable in case the target is offline or supporting fallback navigation via Firefox (tunneled via an appropriate VPN service) in case of persistent congestion of the TOR network (\market\ is reachable from the surface web). 

\smallskip
\noindent \textit{Limitations.}
Profiles sold immediately after appearing on the market may not be captured by our crawlers. We mitigate this problem by employing parallel crawlers, keeping the crawling time at a minimum. Running our crawling during the (east) European night further decreases the chances that profiles will both appear \textit{and} disappear within our window, albeit the market is not reserved for East European customers only. Further, we cannot assure that a profile `permanent' disappearance may not be due to causes other than a sale. However, the presence of many old, unsold profiles~\cite{campobasso2020impersonation} makes this unlikely.
Similarly, we cannot verify that purchases are not performed by actors other than attackers, for example, researchers or LE. However, given the size of the market and the measured sale trends, it is unlikely that volumes of purchases for `legitimate' purposes affect the overall analysis. Our simulations implicitly assume that the \market\ backend for the data harvesting is independent of the market frontend from which we fetch results; further, we cannot explicitly model the effect of market downtime on sales in our model.





\smallskip
\textbf{Conclusions.}
In this paper we presented a unique data collection and rigorous analysis of data on attackers' profile acquisition on a prominent, still active, cybercrime market for user impersonation at scale. The proposed methodology identifies and addresses general challenges inherent to the problem of monitoring (prominent) criminal underground communities. We reconstruct attacker preferences, profile acquisition trends and sale volumes, and estimate the overall market revenue. We discuss implications of our work by integrating the risk model proposed by Woods and Böhme in \cite{woods2021sok}. 

\smallskip
\textbf{Acknoledgements.}
This work is supported by the ITEA3 programme through the DEFRAUDIfy project funded by Rijksdienst voor Ondernemend Nederland, Grant No. ITEA191010, and by the INTERSCT project, Grant No. NWA.1162.18.301, funded by Netherlands Organisation for Scientific Research (NWO).


\vspace{-2mm}
\bibliographystyle{acm}
\bibliography{acmart}

\appendix

\section*{Appendix}

\vspace{-1mm}
\subsection*{Time window selection}

Tab.~\ref{tab:available-data}
\begin{table}
\begin{center}
\caption{Available data points across monitoring periods.}

\label{tab:available-data}
\small
\scalebox{0.9}{
\begin{tabular}{l r r r}
\toprule
 & \multicolumn{3}{c}{Available data points}\\
 \cmidrule{2-4}
data $\forall d \in D$ & obs days (\%) & profiles (\%) & sales (\%) \\
\midrule 
$L^d_0 \bigcup L^d_{1\cap\ldots\cap6}$  & $107$ ($100.0\%$) & $12'149$ ($100.0\%$) & $2'051$ ($100.0\%$) \\
\midrule
$L^d_0 \cap L^d_{1}$   & $101$ ($94.4\%$) & $11'357$ ($93.5\%$) & $1'193$ ($58.2\%$) \\
$\ldots \cap L^d_{2}$     & $89$ ($83.2\%$) & $9'778$ ($80.5\%$) & $1'423$ ($69.4\%$) \\
$\ldots \cap L^d_{3}$     & $86$ ($80.4\%$) & $9'445$ ($77.7\%$) & $1'593$ ($77.7\%$) \\
$\ldots \cap L^d_{4}$     & $77$ ($72.0\%$) & $8'071$ ($66.4\%$) & $1'501$ ($73.2\%$) \\
$\ldots \cap L^d_{5}$     & $72$ ($67.3\%$) & $7'560$ ($62.2\%$) & $1'520$ ($74.1\%$) \\
$\ldots \cap L^d_{6}$     & $67$ ($62.6\%$) & $6'860$ ($56.5\%$) & $1'432$ ($69.8\%$) \\
\bottomrule
\end{tabular}}
\end{center}
\vspace{-6mm}
\end{table}
reports the relative fraction of overall observations that can be derived up to each monitoring day $n$. We note that $L^d_0 \bigcup L^d_{1\cap\ldots\cap6}$ (short for $L^d_0 \cup (L^d_0 \cap L^d_1) \cup \ldots \cup (L^d_0 \cap L^d_1 \cap \ldots\cap L^d_6)$) would allow us to maximize the number of appeared profiles as well as observations of sales while satisfying modelling constraints (Sec.~\ref{subsubsec:meth_pred-and-simulation}). However, the sales model would have to account for all the variability in available alternatives at the purchase decision for any day $d$ (Sec.~\ref{subsubsec:meth_pred-and-simulation}), which proved to be computationally unfeasible for $n>2$ (the model fails to converge). $L^d_0 \bigcup L^d_{1\cap\ldots\cap6}$ does, however, represent the overall empirical evidence we have of actual profile appearances and sales; hence, use it as a benchmark to evaluate the trade-off between the fraction of available profiles and the fraction of remaining sales up to observation day $n$.

\vspace{-2mm}
\subsection*{Distinguishing sales from reservations}
\label{subsec:disc_market_claims}

To verify the market's claims stating that the presence of a product exclusively depends on sales and that there are no other stochastic processes involved in the re-appearing of profiles besides the race condition between our crawling and a profile becoming reserved, we check for every listing day $L^d_0$ if a disappeared product reappears in subsequent listing days $L^d\textsubscript{1..6}$ and how often.
Under the assumption that profiles only disappear if they're sold, $L^d\textsubscript{n+1}$ shall always contain a subset of $L^d_n$; the reservation mechanism introduces violations of this hypothesis, so we measure how often it occurs to evaluate how it compares to our expectations and to understand if it poses concerns on the validity of the sales detection technique. We check $L^d\textsubscript{n+1} \cap L^d\textsubscript{n}, \ \forall n \in [1..5]$. By analyzing the dataset containing information about $L^d\textsubscript{0..6}$, out of the $6'860$ profiles available, only $74$ reappeared over the next monitored days, representing the $1.08\%$ of the total, against the expected $2.08\%$\footnote{Given the maximum duration of a reservation being $30$ minutes, the probability of missing a reserved profile is at most $2.08\%$, if it remains unsold.}. Further, we do not identify any profile that disappeared for more than $1$ day; these results seem compatible with the assumption that no other stochastic processes are involved in this phenomenon. As we consider $L^d\textsubscript{0,1}$, we can identify false positives introduced when labelling a profile as ``sold'' in the case of a profile reappearing on $L^d\textsubscript{2}$; we identify 23 profiles of this type and correct their label accordingly. 

\subsection*{Interpreting features against MFA dimensions}

\begin{figure}[t]
    \centering
    \includegraphics[width=0.85\columnwidth,keepaspectratio]{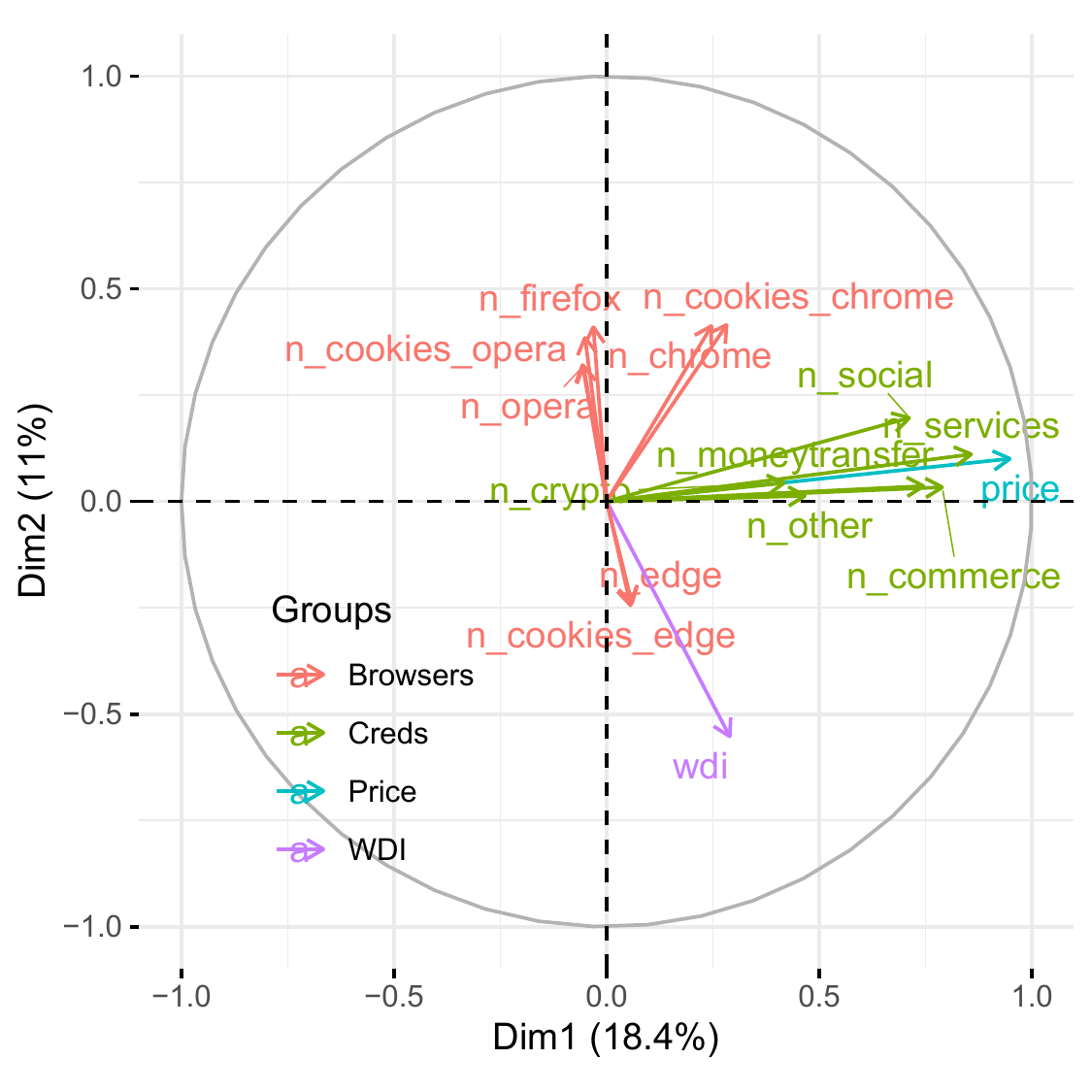}
    \caption{Two-dimensional representation of the variable vector space in output of the MFA.}
    \vspace{-4mm}
    \label{fig:mfa_quanti_var}
\end{figure}

\begin{figure}
    \centering
    \includegraphics[width=0.94\columnwidth,keepaspectratio]{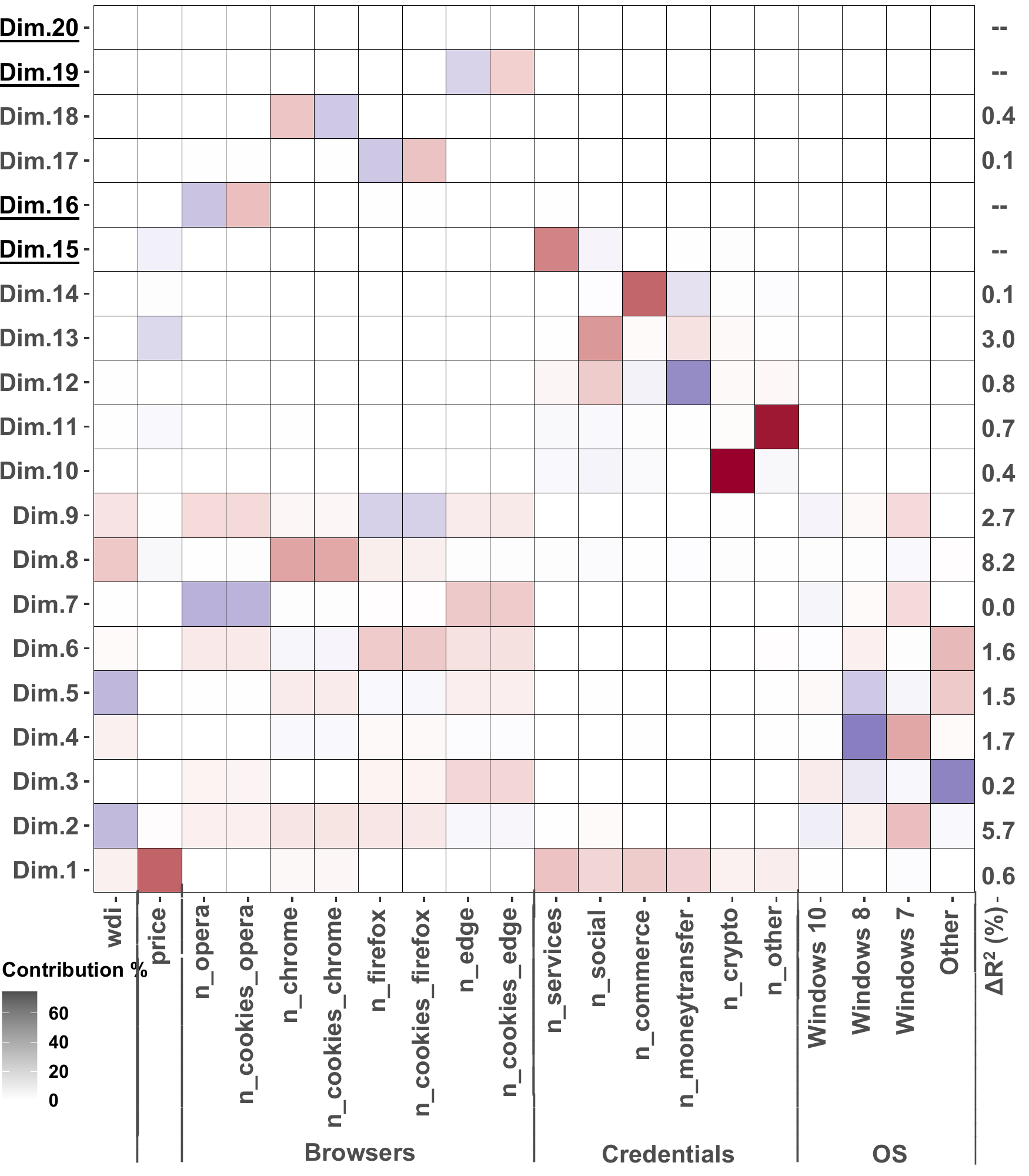}
    \vspace{-2mm}
    \caption{Variables' contribution to MFA dimensions. Underlined dimensions are not included in the final model. For brevity we include here the amount of variance explained by each dimension ($\Delta R^2$) in the final model in the last column of the matrix.}
    \label{fig:mfa_contrib_plot_total}
\end{figure}

Tab.~\ref{tab:mfa_var} reports the original variable variance captured by all MFA dimensions.
Fig.~\ref{fig:mfa_quanti_var}
provides a representation over the two predominant dimensions (\texttt{Dim.1,2}, accounting for $29.44\%$ of the overall data variance) of the (quantitative, as opposed to categorical) variable vector space; the projection of each vector onto each dimension represents how influential that variable is on the dimension, normalized at a group level; 
the closer variables are over a specific dimension, the more that dimension captures correlation among those variables. 
Colors represent variable groups; unsurprisingly, variables within the same group tend to be closely related to each other. \texttt{Browsers} and \texttt{WDI} appear to be of main relevance for \texttt{Dim.2}, whereas \texttt{Dim.1} represents mostly price and available credentials. Price and available resources seem to be highly and positively correlated (in agreement with~\cite{campobasso2020impersonation}); interestingly, \texttt{WDI} is highly but negatively correlated to the browser(s) characteristics of the affected user over these two dimensions; in particular, it emerges that profiles abundant in Edge profiles and cookies originate from more wealthy countries. From Fig.~\ref{fig:mfa_contrib_plot_total}, 
it is possible to observe that dimensions from $10$ to $15$ predominantly characterize profiles in terms of their available credentials; however, due to their low eigenvalues and variance captured, they fail to provide remarkable qualitative insights on the profile construction. Nonetheless, a few considerations can be done. \texttt{Dim.12} shows that generally profiles present an inverse correlation between the number of social and moneytransfer credentials. On the other hand, \texttt{Dim.13} indicates that whenever they are associated, and credentials from social media platforms are predominant, those tend to have a lower price than the average. 
For a more complete perspective on the relations between variables across different dimensions, we provide a repository containing all the possible combinations of two dimensions (as in Fig.~\ref{fig:mfa_quanti_var})\footnote{Link to the resources: \url{https://gitlab.tue.nl/impaas-mfa-plots/impaas-mfa-plots}}. 

\begin{table}[t]
\vspace{-2mm}
\begin{center}
\caption{Captured variance by MFA dimensions.}
\label{tab:mfa_var}
\footnotesize
\scalebox{0.78}{
\begin{tabular}{p{0.1\columnwidth}p{0.06\columnwidth}p{0.06\columnwidth}p{0.06\columnwidth}p{0.06\columnwidth}p{0.06\columnwidth}p{0.06\columnwidth}p{0.06\columnwidth}p{0.06\columnwidth}p{0.06\columnwidth}p{0.08\columnwidth}}
\toprule
\texttt{Dim.} & \texttt{1} & \texttt{2} & \texttt{3} & \texttt{4} & \texttt{5} & \texttt{6} & \texttt{7} & \texttt{8} & \texttt{9} & \texttt{10} \\
\midrule
var (\%)& $18.41$ & $11.02$ & $9.57$ & $9.37$ & $9.08$ & $8.83$ & $8.28$ & $7.37$ & $7.31$ & $2.58$ \\
tot (\%)& $18.41$ & $29.44$ & $39.01$ & $48.38$ & $57.46$ & $66.29$ & $74.57$ & $81.94$ & $89.25$ & $91.83$ \\
\midrule
\texttt{Dim.} & \texttt{11} & \texttt{12} & \texttt{13} & \texttt{14} & \texttt{15} & \texttt{16} & \texttt{17} & \texttt{18} & \texttt{19} & \texttt{20} \\
\midrule
var (\%)& $2.37$ & $1.47$ & $1.23$ & $1.10$& $0.63$ & $0.48$ & $0.37$ & $0.35$ & $0.18$ & $0.00$ \\
tot (\%)& $94.20$ & $95.67$ & $96.90$ & $98.00$ & $98.62$ & $99.10$ & $99.48$ & $99.82$ & $100.00$ & $100.00$ \\
\bottomrule
\end{tabular}}
\end{center}
\vspace{-6mm}
\end{table}

\subsection*{Model evaluation}

\label{sec:appendix-model}

We report the coefficients for the full model in the table below. In parenthesis, the standard error for each added dimension.
\begin{center}
\scalebox{0.74}{
\small
\begin{tabular}{p{0.1\columnwidth}p{0.09\columnwidth}p{0.1\columnwidth}p{0.09\columnwidth}p{0.1\columnwidth}p{0.1\columnwidth}p{0.09\columnwidth}p{0.1\columnwidth}p{0.12\columnwidth}}
\toprule
$\beta_0$     & \texttt{Dim.8} & \texttt{Dim.2} & \texttt{Dim.13} & \texttt{Dim.9} & \texttt{Dim.4} & \texttt{Dim.6} & \texttt{Dim.5} & \texttt{Dim.12} \\
\midrule
$-2.51^{***}$ & $0.62^{***}$   & $-0.41^{***}$  & $1.02^{***}$    & $0.32^{***}$   & $0.19^{***}$   & $0.38^{***}$   & $-0.17^{***}$  & $-0.52^{***}$   \\
$(0.05)$      & $(0.04)$       & $(0.04)$       & $(0.09)$        & $(0.04)$       & $(0.03)$       & $(0.04)$       & $(0.04)$       & $(0.08)$ \\
\midrule
\end{tabular}}
\scalebox{0.74}{
\small
\begin{tabular}{p{0.1\columnwidth}p{0.09\columnwidth}p{0.1\columnwidth}p{0.09\columnwidth}p{0.1\columnwidth}p{0.1\columnwidth}p{0.09\columnwidth}p{0.1\columnwidth}p{0.12\columnwidth}}
              & \texttt{Dim.11} & \texttt{Dim.1} & \texttt{Dim.10} & \texttt{Dim.18} & \texttt{Dim.3} & \texttt{Dim.14} & \texttt{Dim.17} & \texttt{Dim.7} \\
\midrule
              & $0.44^{***}$    & $0.06^{*}$     & $0.28^{***}$    & $-0.76^{***}$   & $-0.35^{***}$  & $-0.30^{**}$    & $0.38^{*}$      & $0.09^{*}$    \\
              & $(0.06)$        & $(0.02)$       & $(0.05)$        & $(0.18)$        & $(0.03)$       & $(0.10)$        & $(0.17)$        & $(0.04)$      \\
\bottomrule
\multicolumn{9}{l}{\footnotesize $std(c|day)=0.25$, AIC=$6564.8$, BIC=$6696.9$, $R\textsuperscript{2}m=0.264$, $R\textsuperscript{2}c=0.278$, \# Obs=$11'357$,} \\
\multicolumn{9}{l}{\footnotesize $^{***}p<0.001$, $^{**}p<0.01$, $^{*}p<0.05$} \\
\end{tabular}
}
\end{center}

We proceeded to build our final model adding one variable at a time and performing the analysis of the variance (ANOVA) for each new model. 
The final model explains the $27.8\%$ of the data theoretical variance.

\vspace{-2mm}
\paragraph*{Model performance.} 

In Sec.~\ref{subsec:disc_model_sales}, we report model performance in terms of $R^2$; $R^2_m$ represents the fraction of variance explained by the fixed effects; $R^2_c$ includes variance explained by both fixed and random effects. $std(c|day)$ is the standard deviation of the random effect at the intercept.

\begin{figure}
    \centering
    \includegraphics[width=0.62\columnwidth,keepaspectratio]{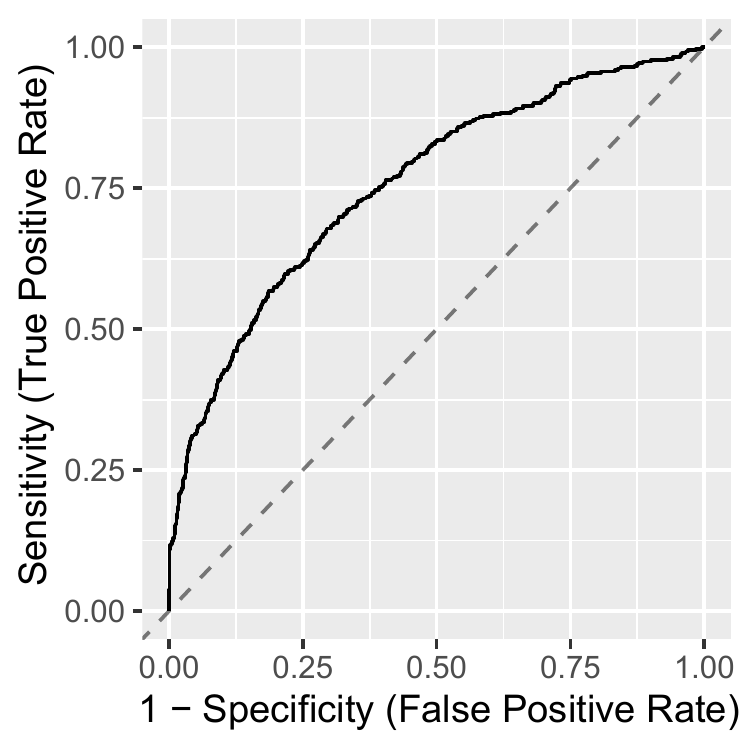}
    \caption{ROC curve with median AUC (0.757) over $1'000$ simulations.}
    \label{fig:roc_median}
    \vspace{-3mm}
\end{figure}

We also compare the adopted model (accounting for purchase alternatives on a given day) to the same fixed effect model over the dataset with all the sales (regardless of how reliable the data collection was up to day $n$, first row of the same table). This results in a fitted model with same coefficient directionality, but achieving only an $R^2=0.14$ against the obtained $R^2 = 0.278$ of the regression accounting for daily profile clusters for purchase alternatives. This further corroborates the importance of modelling the sales process with (tractable) factors accounting for the stochasticity introduced by alternative options on the purchase decision.
To evaluate the discriminatory power of our sales prediction model, we run $1'000$ simulations to cross-validate our model with a randomly selected training set accounting for $2/3$ of the full dataset and validate it with the remaining records. For each simulation we calculate the related area under curve (AUC). The median AUC value amounts to $0.757$, and the related ROC curve is reported in Fig.~\ref{fig:roc_median}; the performance of the model is stable across simulations ($68.2\%$ $CI=[0.746, 0.768]$).




\subsection*{Data reconstruction and simulation} 
\label{app:data_reconstr_sim}

As mentioned in Sec.~\ref{subsec:res_dataprep}, data collected from January $21\textsuperscript{st}$ to June $30\textsuperscript{th}$ 2021 spans over a period of $161$ days; the considered dataset $L^d_\textsubscript{0,1}$ offers $101$ days with complete observations, leaving $60$ days \textit{d} to recreate or simulate as follows: (a) $6$ days \textit{d} have $L^d_0$ and no $L^d_1$; (b) $42$ do not have $L^d_0$, but have $L^d_1$, (c) $6$ have $L^d_2$, $1$ has $L^d_3$, while (d) $4$ have the last $24$ hours market recap.
For one day only (Feb 14\textsuperscript{th}, 2021) we have no information at all; by looking at expected or collected profiles in the surrounding days (from 13 to 16 Feb), it appears that there was only $1$ profile listed, thus leading us to conclude that for that day we would not have captured anything regardless. As mentioned in Sec.~\ref{subsubsec:meth_feature_extract}, we predict the sale outcome for the profiles in (a) using two different cutoff values for a more rigorous evaluation, one calculated to minimize false negatives (``stringent cutoff'' - spec: $0.95$, sens: $0.304$), and another more ``generous'' to improve the prediction's sensitivity (spec: $0.80$, sens: $0.604$); respectively, predictions report $53/792$ sold profiles in the former case and $176/792$ in the latter. To simulate products in (b), we compute the expected number of profiles by looking at the ratio of profiles counted during $L^d_1$ and the offered $L^d_0$ ($17.6\%$, which is below the expected $25\%$ due to \LIM{4}); in case we miss $L^d_1$ (c), we rely on the first $L^d_i$ available by adjusting the previously calculated ratio with an additive factor approximating sales until that day, calculated from dataset $L^d\textsubscript{0..6}$. Finally, to simulate products in (d), we calculate the average expected products as the ratio between the days for which we have available both market report and $L^d_0$ profiles ($16.20\%$ of reported profiles are collected during $L^d_0$ based on $95$ observations; the low percentage is caused again from \LIM{4}). With this information, we can now simulate the listings: we simulate the full market $10'000$ times by sampling with replacement an inversely proportional number of profiles from the $3$ right-most and $3$ left-most days for which $L^d_0$ is available, including days from (a), and we do the whole process twice using the two different threshold values for (a).

\begin{figure}[!t]
    \centering
    \includegraphics[width=0.86\columnwidth,keepaspectratio]{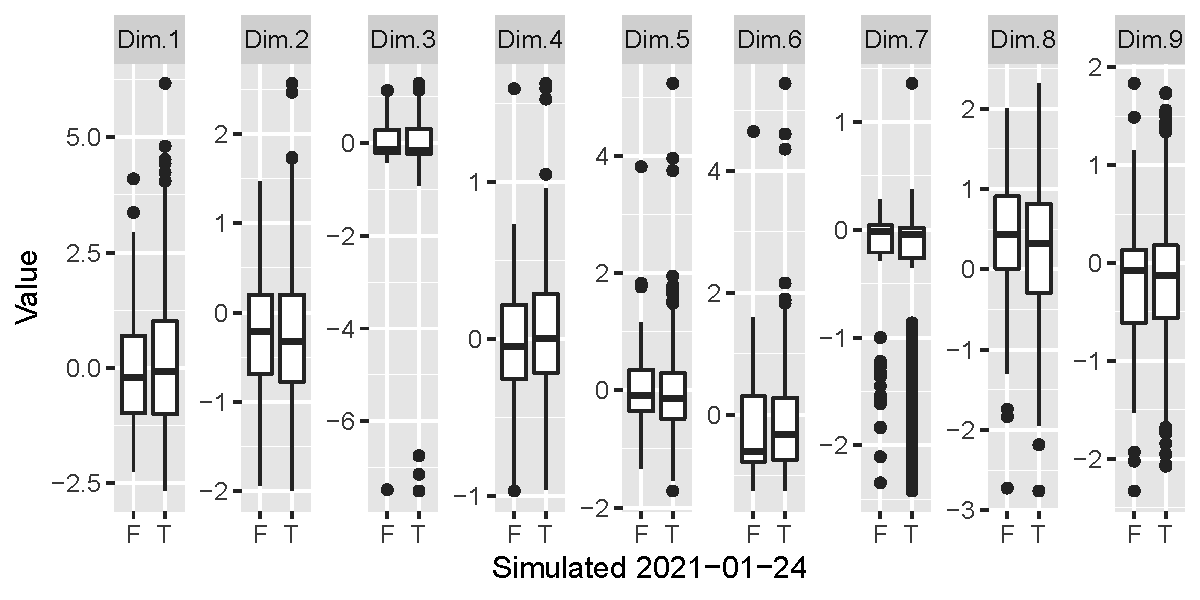}
    \vspace{-5mm}
    \caption{Simulation for January 24\textsuperscript{th}, all neighbors.}
    \label{fig:full0124}
\end{figure}
\begin{figure}[!t]
    \centering
    \includegraphics[width=0.86\columnwidth,keepaspectratio]{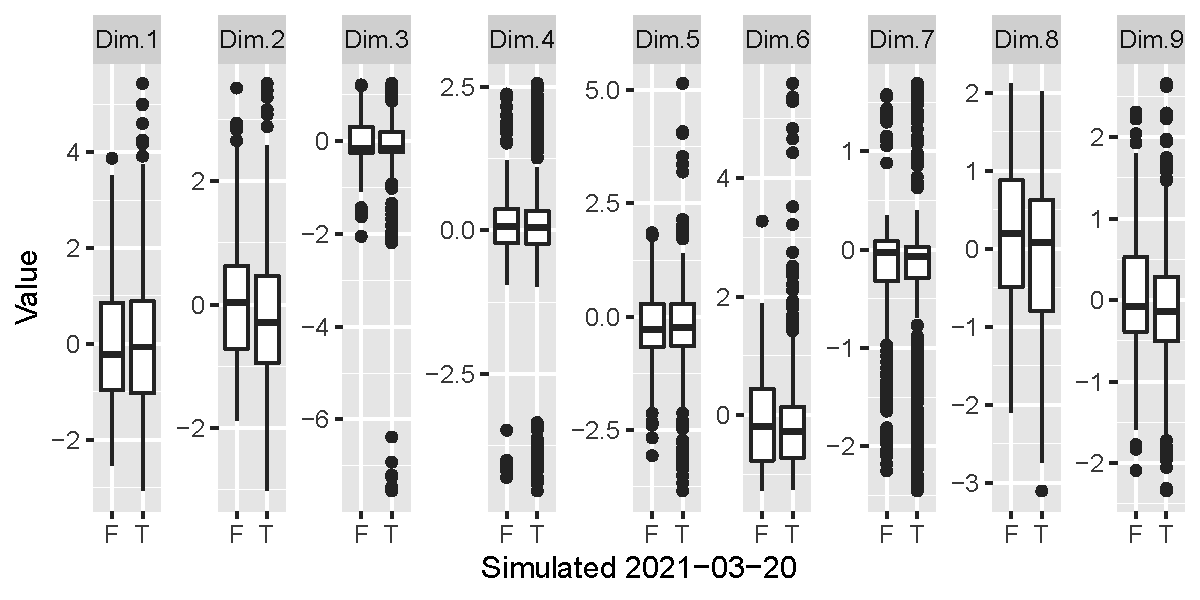}
    \vspace{-5mm}
    \caption{Simulation for March 20\textsuperscript{th}, all neighbors.}
    \label{fig:full0320}
    \vspace{-2mm}
\end{figure}

\begin{figure}[!t]
    \centering
    \includegraphics[width=0.86\columnwidth,keepaspectratio]{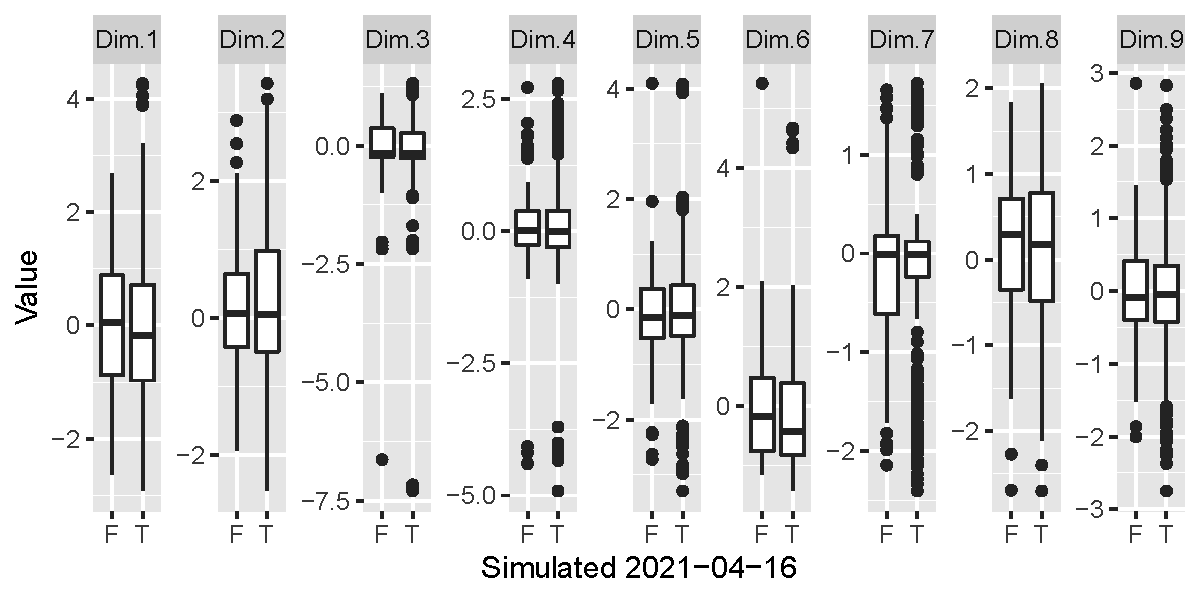}
    \vspace{-5mm}
    \caption{Simulation for April 16\textsuperscript{th}, partial neighbors.}
    \label{fig:partial0416}
\end{figure}

\begin{figure}[!t]
    \centering
    \includegraphics[width=0.86\columnwidth,keepaspectratio]{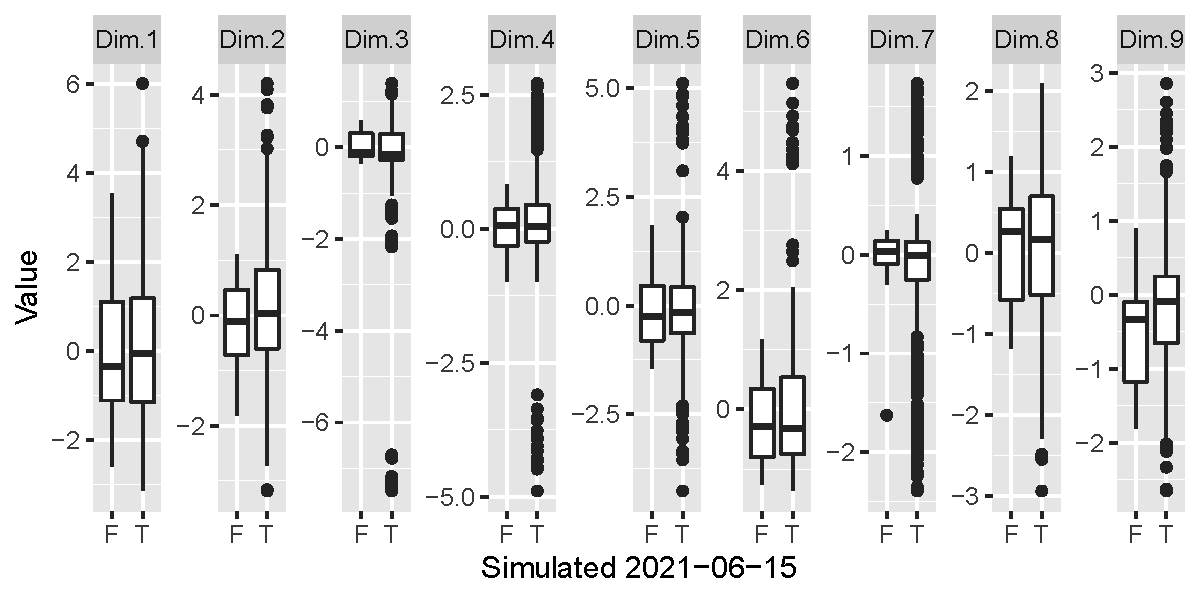}
    \vspace{-5mm}
    \caption{Simulation for June 15\textsuperscript{th}, partial neighbors.}
    \label{fig:partial0615}
    \vspace{-2mm}
\end{figure}

\vspace{-4mm}
\paragraph*{Simulation validation.}
Our simulation strategy assumes that profiles appearing on the market on a certain day are similar to those appeared immediately before or after that day. To verify this assumption, we extract with replacement from the $six$ closest $L_0$ listing days the total expected profiles; the extraction process is weighted based on how close a listing day is to the target listing day to refill. We simulate $1'000$ times the product listings of days for which we already have complete information, and we compare the MFA dimensions for the simulated profiles to the actual profiles. To visualize, we select four random $L_0$, two for which have all neighbor days with full information (e.g., Jan $24\textsuperscript{th}$ is the $L_0$ to simulate and we have all $L_0$ from Jan $21\textsuperscript{st}$ to Jan $27\textsuperscript{th}$), and two for which we do not (e.g., the simulation extracts profiles from the six closest $L_0$ which may be `further' than three days distance from the $L_0$ to simulate). The results are reported in Figures~\ref{fig:full0124}, \ref{fig:full0320}, \ref{fig:partial0416} and \ref{fig:partial0615}: for each dimension, the left boxplot shows the dimensions of the actual profiles, while the right one shows those from simulated profiles. The figures suggest that distributions across different dimensions show no significant differences between the expected and simulated values, suggesting that profiles appearing over contiguous days are similar to each other. A set of Wilcoxon Sign-ranked tests confirms this observation for the observed days across all dimensions used in the model in the $86\%$ of cases, implying that our simulation strategy can reproduce a similar distribution of profiles for the days for which we have no observation.



\subsection*{Round up factor for market volumes} When estimating the market size and revenue, we have to account that (a) we consider only sales up to one day of market activity and (b) we sample only the $25\%$ of the available products at Moscow's midnight. In Sec.~\ref{subsubsec:res_data_diag}, we discussed about the dimensions similarity between products sold within $24$ hours and those sold later; to estimate sales happening after $L^d_1$, we consider $L^d\textsubscript{0..6}$ and compute the fraction of observed sold profiles during $L^d\textsubscript{2..6}$ over the whole period, accounting for the $49\%$ of all sales that would occur during the full six days period of monitoring. Therefore, we'll adjust our sales estimation to cover this fraction of unmeasured sales. With regards to (b), we empirically observed that the listing of a profile can be delayed (up to) some hours; comparing $L^d_0$ and $L^d_1$ cardinality, we note that we sample $17.58\%$ of products on average, instead of the expected $25\%$. These two factors scale our estimations of $11.14$ to represent the actual volumes of \market. To err on the conservative side and to aid comparisons, we round it down to 10 in the paper presentation.

\end{document}